\documentclass{article}

\usepackage[final]{corl_2020} 

\usepackage{hyperref}
\usepackage{url}

\usepackage[utf8]{inputenc} 
\usepackage[T1]{fontenc}    
\usepackage{hyperref}       
\usepackage{url}            
\usepackage{booktabs}       
\usepackage{amsfonts}       
\usepackage{nicefrac}       
\usepackage{microtype}      
\usepackage{algorithmic}
\usepackage{balance}
\usepackage{amsmath}
\usepackage{amssymb}
\usepackage{amsthm}
\usepackage{mathrsfs}
\usepackage{subfigure}
\usepackage{graphicx}
\usepackage{caption}
\usepackage{tabularx}
\usepackage[ruled,linesnumbered]{algorithm2e}
\usepackage{algorithmic}
\usepackage[nolist]{acronym}
\usepackage{bbm}
\usepackage{wrapfig}
\usepackage{float}
\usepackage{siunitx}

\usepackage{enumitem}

\DeclareMathOperator{\tr}{\rm{tr}}

\newcommand{\rd}{{\mathrm d}}

\newcommand{\calU}{{\cal{U}}}

\newcommand{\vx}{{\bf x}}

\newcommand{\vf}{{\bf f}}

\newcommand{\vG}{{\bf G}}
\newcommand{\vu}{{\bf u}}

\newcommand{\vw}{{\bf w}}

\newcommand{\vR}{{\bf R}}

\newcommand{\vSigma}{{\mbox{\boldmath$\Sigma$}}}

\newcommand{\argmin}{\operatornamewithlimits{argmin}}
\newcommand{\T}{^\mathrm{T}}

\newcommand{\Qb}{\mathbb{Q}}

\newcommand{\Rb}{\mathbb{R}}
\newcommand{\Eb}{\mathbb{E}}

\begin{acronym}
\acro{ML}{Machine Learning}
\acro{DNN}{Deep Neural Network}
\acro{DP}{Dynamic Programming}
\acro{FBSDE}{Forward-Backward Stochastic Differential Equation}
\acro{FSDE}{Forward Stochastic Differential Equation}
\acro{BSDE}{Backward Stochastic Differential Equation}
\acro{LSTM}{Long-Short Term Memory}
\acro{FC}{Fully Connected}
\acro{DDP}{Differential Dynamic Programming}
\acro{HJB}{Hamilton-Jacobi-Bellman}
\acro{PDE}{Partial Differential Equation}
\acro{PI}{Path Integral}
\acro{NN}{Neural Network}
\acro{GPs}{Gaussian Processes}
\acro{SOC}{Stochastic Optimal Control}
\acro{RL}{Reinforcement Learning}
\acro{MPOC}{Model Predictive Optimal Control}
\acro{IL}{Imitation Learning}
\acro{RNN}{Recurrent Neural Network}
\acro{FNN}{Feed-forward Neural Network}
\acro{Single FNN}{Single Feed-forward Neural Network}
\acro{DL}{Deep Learning}
\acro{SGD}{Stochastic Gradient Descent}
\acro{SDE}{Stochastic Differential Equation}
\acro{2BSDE}{second-order BSDE}
\acro{CBF}{Control Barrier Function}
\acro{QP}{Quadratic Program}
\acro{CBF}{Control Barrier Function}
\acro{PDIPM}{Primal-Dual Interior Point Method}
\end{acronym}
\newtheorem{thm}{Theorem}

\title{Safe Optimal Control Using Stochastic Barrier Functions and Deep Forward-Backward SDEs}

\author{
  Marcus A. Pereira\\
  Institute for Robotics and Intelligent Machines (IRIM)\\
  Georgia Institute of Technology, \\
  Atlanta, GA, 30332\\
  \texttt{mpereira30@gatech.edu} \\
  \And
  Ziyi Wang \\
  School of Aerospace Engineering \\
  Georgia Institute of Technology, \\
  Atlanta, GA, 30332 \\
  \texttt{ziyiwang@gatech.edu} \\
  \AND
  Ioannis Exarchos \\
  Department of Computer Science\\
  Stanford University, Stanford, CA 94305 \\
  \texttt{exarchos@stanford.edu} \\
  \And
  Evangelos A. Theodorou\\
  School of Aerospace Engineering\\
  Georgia Institute of Technology, \\
  Atlanta, GA, 30332\\
  \texttt{evangelos.theodorou@gatech.edu} \\
}

\begin{document}
\maketitle

\setlength{\intextsep}{4pt} 
\setlength{\floatsep}{4pt} 
\setlength{\textfloatsep}{2pt} 

\begin{abstract}
  This paper introduces a new formulation for stochastic optimal control and stochastic  dynamic  optimization that ensures safety with respect to state and control constraints.  The proposed  methodology brings together concepts such as  Forward-Backward Stochastic Differential Equations,  Stochastic  Barrier Functions, Differentiable Convex Optimization and Deep Learning.  Using the aforementioned concepts, a Neural Network architecture is designed for safe trajectory optimization in which learning can be performed in  an end-to-end fashion. Simulations are performed on three systems to show the efficacy of the proposed methodology. 
\end{abstract}



\section{Introduction}
\par Recent advancements in the areas of stochastic optimal control theory and machine learning create new opportunities towards the development of scalable algorithms for stochastic dynamic optimization. Despite the progress, there is a scarcity of methodologies that have the characteristics of being  solidly grounded in first principles, have the flexibility of deep learning algorithms in terms of representational power, and can be deployed on systems operating in safety critical scenarios.  
\par Safety plays a major role in designing any engineering system in various industries ranging from automobiles and aviation to energy and medicine. With the rapid emergence of various advanced autonomous systems, the control system community has investigated various techniques such as barrier  methods \cite{nguyen2016exponential}, reachable sets \cite{althoffACM2011}, and discrete approximation \cite{mitra2013} to ensure safety certifications.  However with the recent introduction of \acp{CBF} \cite{ames2014control,ames2016control,ames2019control}, there has been a growing research interest in the community in designing controllers with verifiable safety bounds.
\par \acp{CBF} provide a measure of safety to a system given its current state. As the system approaches the boundaries of its safe operating region, the \acp{CBF} tend to infinity, leading to the name "barrier". \cite{nguyen2016exponential}, \cite{ames2019control} have implemented \acp{CBF} for deterministic systems in robotics such as bi-pedal walking on stepping stones. However, literature on \acp{CBF} for systems with stochastic disturbances is very scarce. Very recently, \cite{clark2019control} introduced stochastic \acp{CBF} for relative degree 1 barrier functions which means the function that defines the safe set, depends only on the states that are directly actuated. While the concept of stochastic \acp{CBF} is fairly recent,  \cite{clark2019control} only demonstrated their applicability to very simplified stochastic systems in conjunction with control Lyapunov functions. The idea of merging the concepts of CBFs with Control Lyapunov functions leads to some interesting results \cite{ames2014control,ames2016control,ames2019control} in sense of stability of the system. 

\par We here propose to explore the inclusion of the concept of stochastic \acp{CBF} within the \ac{SOC} framework which essentially leads to the problem of solving the \ac{HJB} equation on a constrained solution set. Solving the \ac{HJB} amounts to overcoming the curse of dimensionality. Popular solution methods in literature include iLQG \cite{todorov2005generalized}, Path-Integral Control \cite{theodorou2010generalized} and the \acp{FBSDE} framework \cite{exarchos2018stochastic} which tackle the \ac{HJB} through locally optimal solutions. Of these, the algorithms based on \acp{FBSDE} are the most general in the sense that they neither require assumptions to simplify the \ac{HJB} \ac{PDE} nor do they require Taylor's approximations of the dynamics and value function \cite{todorov2005generalized}. More recently, with the introduction of deep learning methods to solve high-dimensional parabolic \acp{PDE} \cite{han2018solving}, deep learning based solutions of the \ac{HJB} \ac{PDE} using so called Deep \ac{FBSDE} controllers have emerged \cite{pereira2019learning, wang2019deep, wang2019deep_control_multi}. These algorithms leverage importance sampling using Girsanov's theorem of change of measure \cite[Chapter 5]{shreve2004stochastic} for \acp{FBSDE} within deep learning models for sufficient exploration of the solution space.  Additionally, they have been shown to scale to high-dimensional linear systems \cite{pereira2020feynman} and highly nonlinear systems for finite-horizon \ac{SOC}.
\par Paralleling the work on deep learning-based \ac{SOC} is deep learning-based optimization layers, which aims to increase the representational power of \ac{DL} models. In \cite{amos2017optnet}, the differentiable optimization layer was introduced to explicitly learn nonlinear mappings characterized by a \ac{QP}. Additionally, they introduce an efficient approach to compute gradients for backpropagation through an optimization layer using the KKT conditions which allows such layers to be incorporated within Deep \acp{FBSDE} which requires differentiability of all its subcomponents. 
\par To the best of our knowledge, literature on combining \ac{SOC} with Stochastic \acp{CBF} and differentiable convex optimization, all embedded within a deep model is non-existent. Additionally this adds a sense of interpretability to the entire framework compared to the relative black box nature of deep neural networks used for end-to-end control. Our contributions are as follows:

\begin{enumerate}
    \item A novel end-to-end differentiable architecture with embedded safety using Deep \acp{FBSDE}. 
    \item Safe-RL algorithm combining differentiable optimization layers and Deep \acp{FBSDE}.
    \item An extension of existing theory on Stochastic Zeroing Control Barrier Functions (ZCBFs).
\end{enumerate}
\par The rest of this paper is organized as follows: Section \ref{problem_formulation} goes over the problem formulation and Section \ref{fbsde} introduces the \ac{FBSDE} framework. In Section \ref{sCBF}, the stochastic \ac{CBF} is presented, while the differentiable \ac{QP} layer is summarized in Section \ref{differentiable_qp}. The algorithm and network architecture are explained in Section \ref{algorithm}. The simulation results are included in Section \ref{simulation}. Finally, we conclude the paper and present some future directions in Section \ref{future_work}.
\section{Mathematical Preliminaries and Problem formulation} \label{problem_formulation}
\par We consider stochastic dynamical systems that are nonlinear in state and affine in control: 
\begin{align}
    \rd\vx(t) &= \big(\vf(\vx(t)) + \vG(\vx(t)) \vu(t) \big)\rd t +  \vSigma(\vx(t)) \rd \vw(t),
\label{eq:SDE}
\end{align}
where $\vw(t)$ is a $n_w$ dimensional standard Brownian motion, 
and $\vx\in\Rb^{n_{x}}$, and $\vu\in \Rb^{n_u}$ 
denote the state and control vectors, respectively. 
The functions $\vf: \Rb^{n_x} \rightarrow \Rb^{n_x}$, $\vG:\Rb^{n_x} \rightarrow \Rb^{n_x \times n_u}$ and $\vSigma: \Rb^{n_x} \rightarrow \Rb^{n_x \times n_w}$ represent the uncontrolled system drift dynamics, the actuator dynamics, and the diffusion matrix. We assume that the state space is divided into a \textit{safe set} $\mathcal{C}$ which consists of all states that are deemed safe for exploration, and its complement $\Rb^{n_x}\setminus \mathcal{C}$ denoting unsafe states.

\subsection{Stochastic Control Barrier Functions}
\label{sCBF}
For the dynamical system defined by \eqref{eq:SDE}, safety is ensured if  $\vx(t)\in \cal C$ for all $t$ where the set $\cal C$ defines the safe region of operation. $\cal C$ is defined by a locally Lipschitz function $h: \Rb ^{n_x}\rightarrow \Rb$ \cite{ames2014control} as: 
\begin{align}
    \mathcal{C} &= \{\vx : h(\vx)\ge 0 \} \label{C1}\\
    \partial \mathcal{C} &= \{\vx : h(\vx) = 0 \} \label{C2}.
\end{align}

The literature of CBFs makes use of two different kinds of control barrier functions \cite{ames2016control}: \textit{reciprocal} CBFs, whose value approaches infinity when $\vx$ approaches the boundary of the safe set $\partial \mathcal{C}$, and \textit{zeroing} CBFs, whose value approaches zero close to the boundary. In this work we will focus on the latter kind. Without loss of generality, the function $h(\vx)$ which defines the safe set, is a zeroing CBF in itself and guarantees safety if for some class $\mathcal{K}$ function\footnote{A class $\mathcal{K}$ function is a continuous, strictly increasing function $\alpha(\cdot)$ such that $\alpha(0)=0$. } $\alpha(\cdot)$ the following inequality holds $\forall t$: 
\begin{equation}\label{CBFineq}
\frac{\partial h}{\partial \vx}\Big(\vf(\vx)+\vG(\vx)\vu\Big)+\frac{1}{2}\tr\Big(\frac{\partial^2 h}{\partial \vx^2}\vSigma(\vx)\vSigma\T(\vx)\Big)\geq -\alpha(h(\vx)).
\end{equation}
Above, the left-hand-side represents the rate of change of $h$ while the right-hand-side provides a bound to that rate which depends on how close $\vx$ is to the boundary; essentially, the closer the state $\vx$ is to the boundary, the slower the value of $h$ can further decrease, implying that $\vx$ may approach the boundary at ever-decreasing speed. The following theorem formalizes the safety guarantees:
\begin{thm}
Let the safe set $\mathcal{C}$ be defined as per equations \eqref{C1}, \eqref{C2} and let the initial condition $\vx(0) \in \mathcal{C}$. Assume there exists a control process $\vu(t')$ such that for all $t'<t$, equation \eqref{CBFineq} is satisfied. Then, for all $t'<t$ the process $\vx(t)$ has remained within the safe set with probability 1, i.e., $\mathbb{P}(\vx(t)\in \mathcal{C}~ \forall t'<t)=1$.
\end{thm}
The proof is a generalization of Theorem 3 in \cite{clark2020control}, and is given in the Appendix. Note that for any given $\vx$, eq. $\eqref{CBFineq}$ essentially imposes a constraint on the values of control $\vu$ one can apply on the system if one wishes to maintain safety ($\vx \in \mathcal{C}$).

\subsection{Stochastic Optimal Control}

The \ac{SOC} problem is formulated in order to minimize the expected cost given as:
\begin{align}
J\big(\vu \big) = \Eb_{\Qb} \Bigg[\int\limits_{t}^T \big(q(\vx) 
+ \frac{1}{2}\vu\T \vR\vu\big) \rd s +\phi\big(\vx(T)\big) \Bigg], 
\label{eq:cost_functional}
\end{align}
subject to the stochastic dynamics given by \eqref{eq:SDE}, and the constraint that trajectories should remain in the safe set $\mathcal{C}$ at all times. Here, $\phi :\Rb^{n_x}\rightarrow\Rb^+$ is the terminal state cost, $q:\Rb^{n_x}\rightarrow\Rb^+$ denotes the running state cost and  $\vR \in \mathbb{S}_+^{n_u}$ where $\mathbb{S}_+^{n_u}$ represents the symmetric positive definite matrix space. Finally, $\Qb$ is the space of trajectories induced by the controlled dynamics in \eqref{eq:SDE}. We assume that all necessary technical requirements \cite{yong1999stochastic} regarding filtered probability space, Lipschitz continuity, regularity and growth conditions to guarantee existence and uniqueness of strong solutions to \eqref{eq:SDE} are met. Letting $\calU([0,T])$ be the space of admissible controls such that $\vx(t)$ remains within $\mathcal{C}$ over a fixed finite time horizon $T\in[0,\infty)$, the value function related to \eqref{eq:cost_functional} is defined as
$$V\big(\vx,t\big)=\inf_{\vu \in \mathcal{U}[0,T]} J\big(\vu)|_{\vx_0=\vx,t_0=t},$$
and using Bellman's principle of optimality, one can derive the HJB PDE, given by 
\begin{align}
    V_t + \inf_{\vu \in \mathcal{U}[0,T]}\bigg [\frac{1}{2}\text{tr} \big(V_{\vx\vx}\Sigma\Sigma^T\big) + V_{\vx}^T \big(\vf(\vx)+\vG(\vx)\vu\big) + q(\vx)+\frac{1}{2}\vu^TR\vu \bigg] = 0, \quad V\big(\vx, T\big) = \phi\big(\vx\big),
    \label{eq:HJB}
\end{align}
where we drop dependencies for brevity, and use subscripts to indicate partial derivatives with respect to time and the state vector. The term inside the infimum operation defines the Hamiltonian: 
$$\mathcal{H}\big(t, \vx, \vu, V_{\vx}, V_{\vx\vx}\Sigma\Sigma^T\big) = \frac{1}{2}\text{tr}\big(V_{\vx\vx}\Sigma\Sigma^T\big) + V_{\vx}^T \big(\vf\big(\vx)+\vG(\vx)\vu\big) + q(\vx)+\frac{1}{2}\vu^TR\vu.$$
Note that in this case the Hamiltonian is quadratic with respect to $\vu$; if the constraint on $\vu$ in order to remain within the safe set $\mathcal{C}$ (eq.~\eqref{CBFineq}) were not present,  the optimal control could be calculated by setting $\partial\mathcal{H}/\partial \vu = 0$, resulting in $\vu^*(t,\vx)=-\vR^{-1}\vG\T V_{\vx}$. However, the CBF inequality constraint prevents such a closed-form solution for $\vu^*$.

\section{Safe Deep FBSDE Control}
\label{fbsde}
\subsection{Deep \ac{FBSDE} formulation}
Using the nonlinear Feynman-Kac lemma, we can establish an equivalence between the \ac{HJB} \ac{PDE} and the following system of \acp{FBSDE} \cite{yong1999stochastic}:
Given that a solution $\vu^*$ to the constrained minimization of $\mathcal{H}$ exists, the unique solution of \eqref{eq:HJB} corresponds to the following system of FBSDEs,
\begin{align}
    \mathrm{d}\vx(t) &= \big(\vf(\vx(t))+\vG(\vx(t))\vu^*(t) \big)\, \text{d}t + \Sigma\big(\vx(t)\big)\, \text{d}\vw_t, \qquad ~~~~~~\vx(0)=\vx_0,  \qquad \quad  (\text{FSDE}) \\ 
    \mathrm{d}V(t) &=  -\big(q(\vx)+ \frac{1}{2}\vu^{*\mathrm{T}}(t)R\vu^*(t)\big)\, \text{d}t + V_{\vx}^\mathrm{T}(t) \Sigma\big(\vx(t) \big)\, \text{d}\vw_t,~~ V(T)=\phi(\vx(T)),~~ (\text{BSDE}) \label{BSDE}\\
    \vu^*(t) &= \underset{\vu}{\arg\min}\, \mathcal{H}(t) = \underset{\vu}{\arg\min}\,\big\{V_{\vx}^\mathrm{T}(t)\vG(\vx(t))\vu + \frac{1}{2}\vu^TR\vu\big\} , 
    ~\qquad \qquad \qquad \quad (\text{QP})\label{minHamilt}\\
    & \quad\mathrm{s.t.}~\frac{\partial h}{\partial \vx}(\vf(\vx(t))+\vG(\vx(t))\vu)+\frac{1}{2}\tr\Big(\frac{\partial^2 h}{\partial \vx^2}\vSigma(\vx(t))\vSigma\T(\vx(t))\Big)\geq -\alpha(h(\vx(t))).\nonumber
\end{align}
Here, $V(t)$ is a shorthand notation for $V(x(t),t)$ and denotes an evaluation of the function $V(\vx,t)$ along a path of $\vx(t)$, thus $V(t)$ is a stochastic process (same for $V_{\vx}(t)$ and $V_{\vx\vx}(t)$). The above equations are time-discretized and \eqref{minHamilt} is a QP that needs to be solved for each time instant. 

In the above system, $\vx(t)$ evolves forward in time (due to its initial condition $\vx(0) = \vx_0$), whereas $V(\vx(t),t)$ evolves backwards in time, due to its terminal condition $\phi(\vx(T))$. However, a simple backward integration of $V(\vx(t),t)$ would result in it depending explicitly on future values of noise (final term in eq.~\eqref{BSDE}), which is not desirable for a non-anticipating process, i.e., a process that does not exploit knowledge on future noise values. To avoid back-propagation of the BSDE, we utilize \ac{DL}. Specifically, incorporating \ac{DL} \cite{pereira2019learning}  allows us to randomly initialize the value function at start time by treating it as a trainable parameter allowing forward propagation of $V$ instead of backward, and use backpropagation to train the initial $V(\vx(0),0)$, along with an approximation of the gradient function $V_{\vx}(\vx,t)$ which appears in the right-hand-side of eq.~\eqref{BSDE} and in the Hamiltonian of eq.~\eqref{minHamilt}. In this work, we utilize the Deep \ac{FBSDE} controller architecture in \cite{pereira2019learning} as our starting point and add a differentiable stochastic \ac{CBF} layer to ensure safety and maintain end-to-end differentiability.

\subsection{Differential Quadratic Programming Layer}
\label{differentiable_qp}
In \cite{amos2017optnet}, the authors proposed a differentiable optimization layer capable of learning nonlinear mappings characterized by \acp{QP} of the form
\begin{align}\label{eq:qp}
    \bar{\vu} & = \argmin_\vu \frac{1}{2} \vu\T Q(\vu_i) \vu + q(\vu_i)\T \vu\\
    \mathrm{s.t.}& \, C(\vu_i)\vu\leq d(\vu_i) \notag 
\end{align}
where, $\vu_i$ is the \ac{QP} layer's input, $\bar{\vu}$ is the fixed point solution of the \ac{QP} problem as well as the output of the \ac{QP} layer, and the \ac{QP} parameters are $Q: \Rb^{n_u} \rightarrow \Rb^{n_u\times n_u}$, $q:\Rb^{n_u} \rightarrow \Rb^{n_u} $, $C:\Rb^{ n_u} \rightarrow \Rb^{n_q\times n_u}, d:\Rb^{ n_u} \rightarrow \Rb^{n_q}$ and $n_q \text{ is the number of inequality constraints}$.
\par Specifically, during the forward pass through the \ac{QP} layer, the optimization problem is solved by running the \ac{PDIPM} \cite[Chapter 14]{nocedal2006numerical} until convergence. \ac{PDIPM} requires a feasible initialization and searches within the constraints. 
\par Since the Deep \ac{FBSDE} network is trained by backpropagating gradients through time, one needs to be able to pass gradients through \acp{QP} solved at each time step. During backpropagation, instead of relying on the auto-differentiation of forward computations, a linear system formulated from the KKT conditions, that $\bar{\vu}$ must satisfy, is used to find the gradient of $\bar{\vu}$ with respect to the \ac{QP} parameters. In our approach, the differentiable \ac{QP} layer is used to solve the QP problem in eq.~\eqref{minHamilt}. To match the \ac{QP} in \eqref{eq:qp}, we set $Q=R$, $q=V_{\vx}^T\vG$, $C=-\frac{\partial h}{\partial \vx} \vG, d=\alpha(h(\vx))+\frac{\partial h}{\partial \vx} \vf + \frac{1}{2}\tr \big(\frac{\partial^2 h}{\partial \vx^2}\vSigma\vSigma\T\big)$. Unlike \cite{amos2017optnet}, the \ac{QP} parameters are not trainable in our case, as they are pre-defined by our problem. 
\begin{figure*}[ht!]
\centering
  \includegraphics[width=\linewidth]{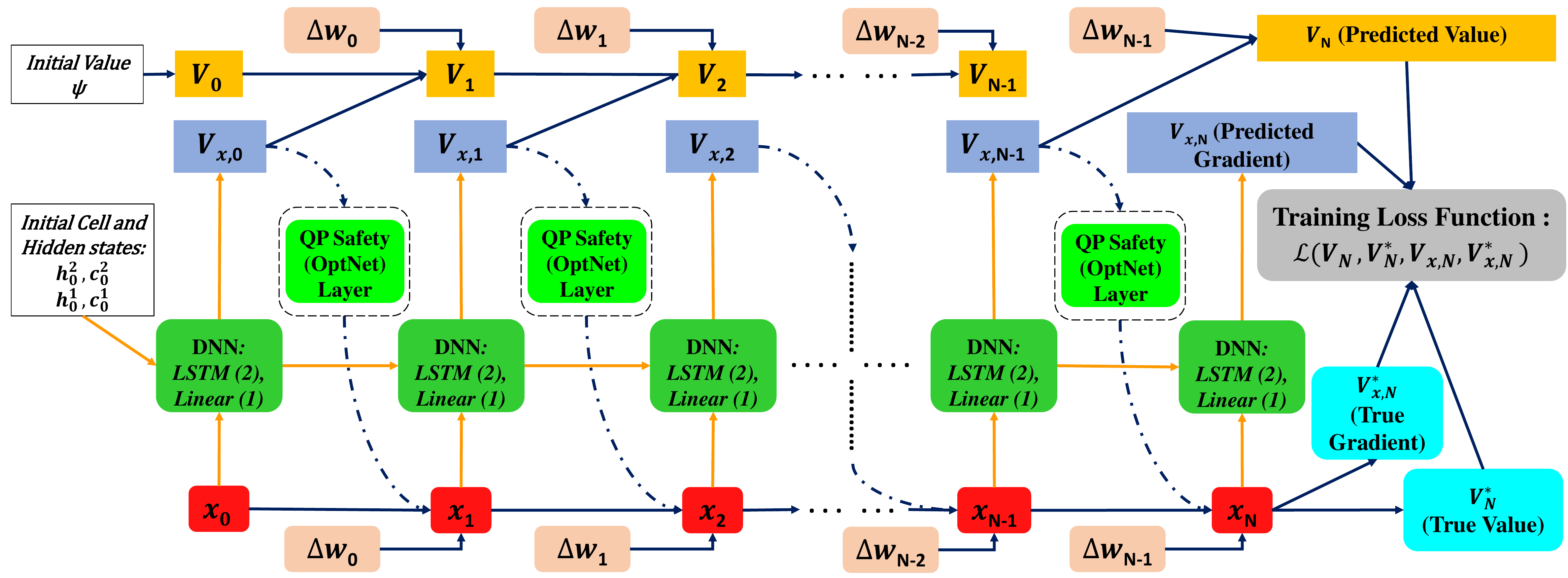}
  \caption{\textbf{Safe \ac{FBSDE} Controller}: Note that the same trainable parameters of the DNN (2 LSTM + 1 Linear) layers are shared across all time steps, and the same noise is used to propagate the value function, $V$ and the system state, $\vx$. Additionally, the initial value, cell and hidden states are trainable.}
\label{fig:FBSDEnetwork}
\end{figure*}
\subsection{Deep FBSDEs for Safe SOC - Algorithm and Network Architecture}
\label{algorithm}
The time discretization scheme is same as that employed in \cite[Section IV]{pereira2019learning}.
The network architecture of the safe \ac{FBSDE} controller is presented in Fig.~\ref{fig:FBSDEnetwork}. The network consists of a \ac{FBSDE} prediction layer (dark green), a safe layer (light green) and two forward propagation processes. The main differences between our proposed network and the one introduced in \cite[Fig. 2]{pereira2019learning} are: 1) the value function gradient at initial time comes from network prediction rather than random initialization, making the predictions more consistent temporally; 2) the propagated and true value function gradients are included in the loss function instead of only using the propagated and true value function. The safe layer, consisting of the differentiable \ac{QP} OptNet layer  \cite{amos2017optnet}, solves the constrained Hamiltonian minimization problem \eqref{minHamilt}. The differentiability of components in the safe layer ensures that the entire network is end-to-end trainable (gradient can flow through all connections in fig. \ref{fig:FBSDEnetwork}).
A summary of the algorithm is given in alg.1 and 2 in the Supplementary Material. 
Given an initial state, the value function gradient at that state is approximated by the \ac{FBSDE} layer prediction and is used to construct the constrained Hamiltonian that is solved by the safe layer, as detailed in alg. 2. 
The safe layer formulates the constrained Hamiltonian minimization as a \ac{QP} and solves for the safe control using the KKT conditions and \ac{PDIPM} interior point solver. For further details of \ac{PDIPM} see \cite{amos2017optnet, mattingley2012cvxgen}. 
Both state $\vx$ and value function $V$ are then propagated forward with the computed safe control. At terminal time $T$, the true value function and its gradient are calculated from the terminal state and compared against their propagated counterparts. The loss function is constructed using the difference between propagated and true value function, the true value function, the difference between propagated and true value function gradient, and the true value function gradient (equation in alg. 1). Training is done using the Adam \cite{kingma2014adam} optimizer.

\section{Simulations}
\label{simulation}
\begin{wrapfigure}{r}{0.3\textwidth}
\vspace{-15 pt}
\centering
\includegraphics[width=0.3\textwidth]{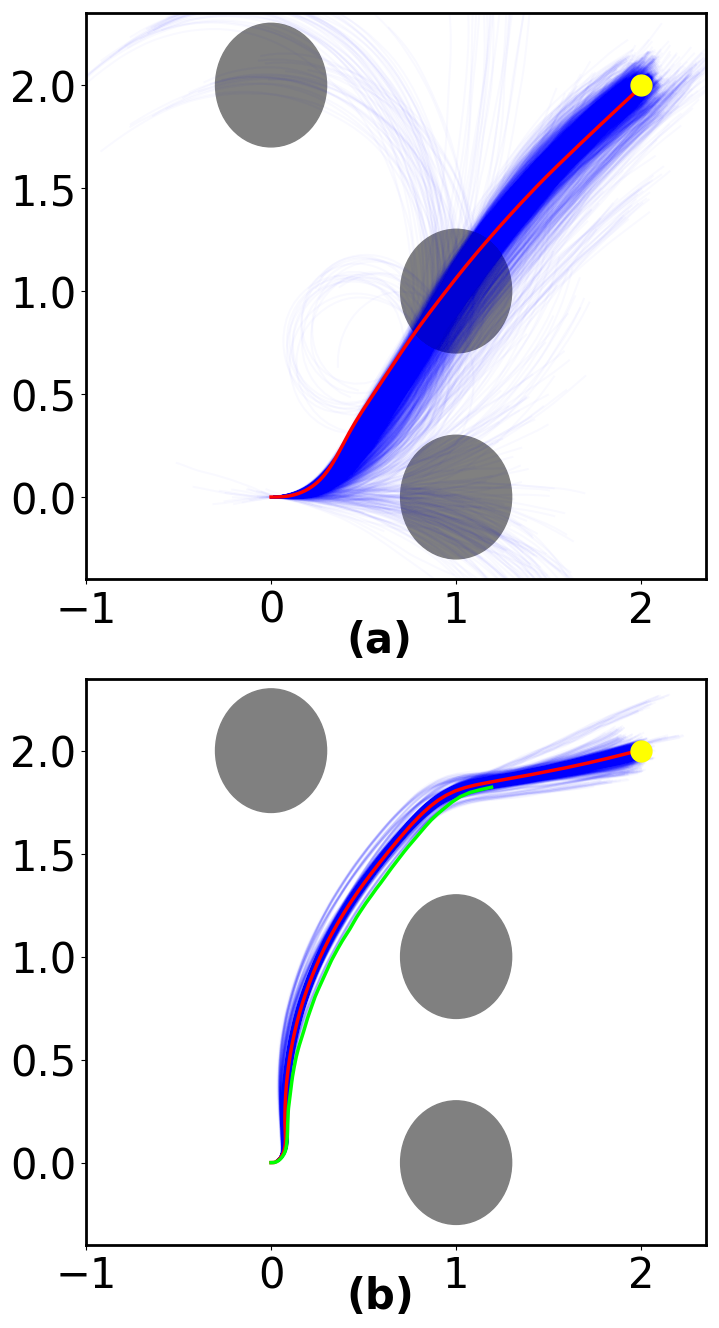}
  \caption{\textbf{Obstacle avoidance:} (a) and (b) represent \textit{unsafe} and \textit{safe} respectively. Policy is learnt avoiding obstacles (\textit{grey}) during training (\textit{blue}) to reach the target (\textit{yellow}). Shown in (\textit{red}) and (\textit{lime}) are mean test and the worst-case trajectories.}
  \vspace{-15 pt}  
\label{fig:2dcar_obstacle}
\end{wrapfigure}
In this section we provide details of the successful application of our proposed Safe \ac{FBSDE} controller on three different systems: pendulum, cart-pole and 2D car in simulation. All simulations are conducted in a safe RL setting which means we begin inside the safe set and learn an optimal controller while never leaving the safe set \textit{during the entire training process}. We have included a comprehensive description of each system in the supplementary material. This includes, the equations of motion written as SDEs in state-space form, the equations of the barrier functions used, the derivations for the gradients, Hessians and Lie-derivatives as required in \eqref{eq:qp} for the Hamiltonian minimization at each time step as well as values for hyperparameters for training the deep Safe FBSDE networks. In all simulations we compare with an \textit{unsafe} FBSDE controller which is work done in \cite{pereira2019learning} without any safety constraints.
The barriers functions used in our simulations take a generic form of,
\begin{equation*}
    h(\vx) = (\text{position constraint}) - \mu(\text{velocity})^2
\end{equation*}
where, $(\text{position constraint})\begin{cases}>0,\quad\forall \vx \in \mathcal{C}\backslash \partial\mathcal{C}\\ =0, \quad \forall \vx \in \partial\mathcal{C}\end{cases}$. The parameter $\mu$ controls \textit{how fast the system can move inside the safe set.} The above formulation causes the barrier function to have a relative degree of 1. For more information see the Supplementary Material.
All the computational graphs were implemented in PyTorch \cite{pytorch} on a system with Intel(R) Xeon(R) CPU E5-1607 v3 @ 3.10GHz, 64GB RAM and a NVIDIA Quadro K5200 GPU. We considered a uniform time discretization of $\Delta t= 0.02$s for all the tasks. 
\begin{figure}[t]
\centering
  \includegraphics[scale=0.4]{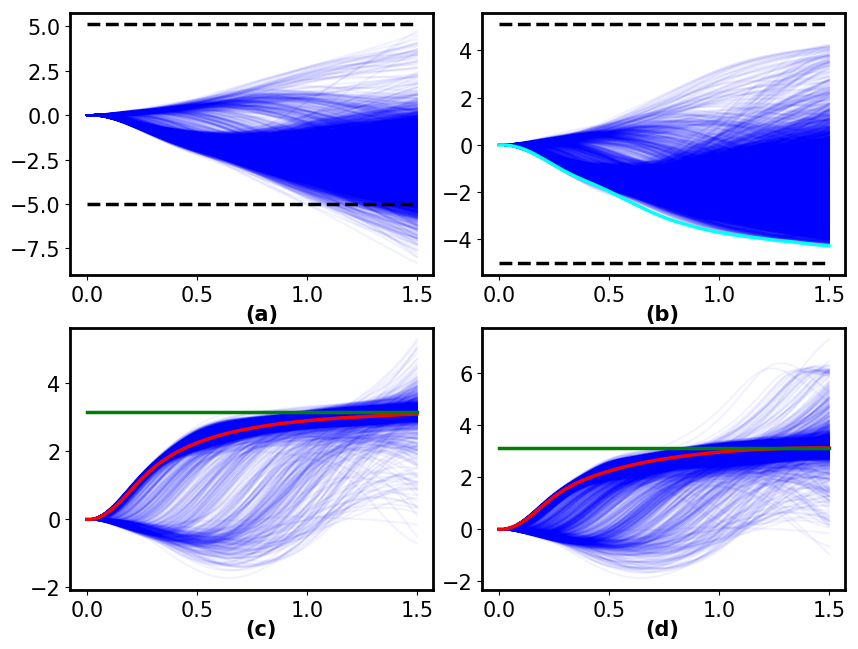}
  \caption{\textbf{Cart-pole Swing-up:} The horizontal axis represents \textit{time (s)} in all subplots while vertical axis represents \textit{cart-position (m)} in (a) and (b) and \textit{pole-angle (rad)} in (c) and (d). Plots (a) and (c) show \textit{unsafe} while (b) and (d) show \textit{safe} trajectories. The safe controller learns to perform a swing-up while always remaining within the bounds (dashed \textit{black} lines) during training (shown by \textit{blue} trajectories). The mean test performance is shown in \textit{red}, target pole-angle is shown in \textit{green} and the worst-case (i.e. closest to either bound during training) is shown in \textit{cyan}.}
\label{fig:cartpole_swingup}
\end{figure}

\begin{figure}[h!]
\centering
  \includegraphics[scale=0.17]{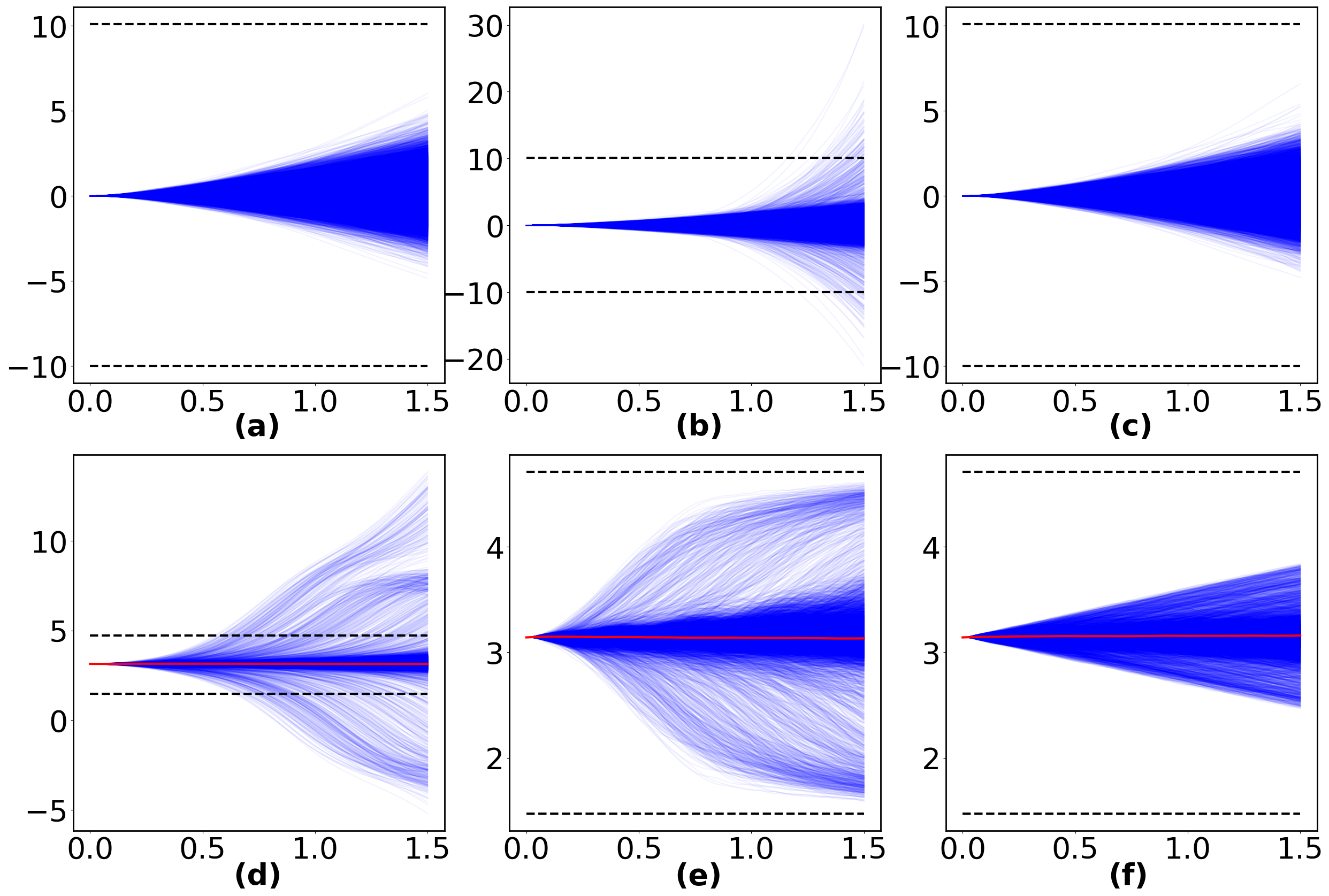}
  \caption{\textbf{Cart-pole Balancing:} The horizontal axis is \textit{time (s)} in all subplots. Upper row plots are \textit{cart-position (m)} and bottom row plots are \textit{pole angular-position (rad)}. Plots (a) and (d) are \textit{unsafe} trajectories, plots (b) and (e) are \textit{one constraint safe} trajectories and plots (c) and (f) are \textit{double constraint safe} trajectories. The safety bounds are indicated by dashed \textit{black} lines and mean performance during testing in shown in \textit{red}. Since all $25,728$ training trajectories are plotted, there is no need to explicitly show the worst-case trajectory.}
\label{fig:cartpole_balancing}
\end{figure}

\subsection{Pendulum Balancing}
\begin{wrapfigure}{r}{0.25\textwidth}
\vspace{-35 pt}
\centering
\includegraphics[width=0.25\textwidth]{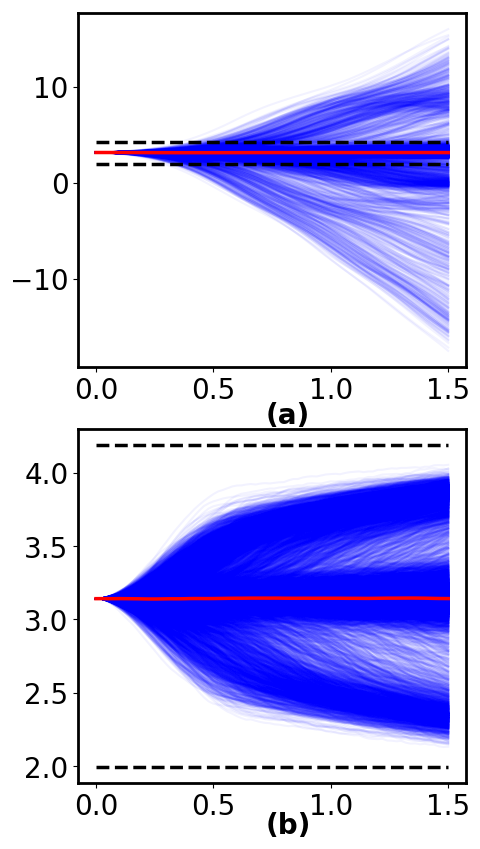}
  \caption{\textbf{Pendulum Balancing:} (a) and (b) represent \textit{safe} and \textit{unsafe} trajectories respectively.}
\vspace{-25 pt}  
\label{fig:pendulum}
\end{wrapfigure}

The state for this system are given by $\vx=[\theta,\,\dot{\theta}]^{\text{T}}$ representing pendulum angle and its velocity. This task requires starting at the unstable equilibrium point of $\vx=[\pi,0]^{\text{T}}$ and balancing around it for a time horizon of $1.5$ seconds. Due to stochasticity, the pendulum is perturbed from this initial position causing it to fall to the stable equilibrium point $\vx=[0,0]^{\text{T}}$. To enforce safety, we impose box constraints on the angular position of the pendulum of $\theta \in [2\pi/3,\, 4\pi/3]$. The results from our simulation are shown in Fig. \ref{fig:pendulum} wherein all the $25,728$ trajectories encountered during training are plotted in \textit{blue} and compared with an unsafe solution. Our proposed controller  Fig. \ref{fig:pendulum}(b) is able to match the unsafe controller  Fig. \ref{fig:pendulum}(a) on average during testing (\textit{red}), while respecting the imposed angular bounds during the entire training process.

\subsection{Cart-pole}The states of the system are given by $\vx=[x_c,\,\theta,\,\dot{x_c},\,\dot{\theta}]^{\text{T}}$
representing the cart-position, pole angular-position, 
cart-velocity and pole angular-velocity respectively. We considered the following 3 different tasks (the time horizon for all tasks is $1.5$ seconds): 

\subsubsection{Swing-up with bounds on cart-position}
This task requires starting from an initial state of $\vx=[0,0,0,0]^{\text{T}}$ and finding a sequence of control actions to reach a final state of $\vx=[0,\pi,0,0]^{\text{T}}$. To enforce safety, we impose constraints on the cart-position of $[-5.0\text{ m},\,5.0\text{ m}]$. The training (\textit{blue}) and average testing (\textit{red}) trajectories from our simulations are shown in Fig. \ref{fig:cartpole_swingup}. Since we cannot plot all $256,128$ training trajectories we randomly sample $10,000$ for plotting. However, we also plot the worst-case trajectory encountered (i.e. trajectory that gets closest to one of the boundaries) to emphasize that no trajectory violates the safety constraint \textit{during the entire training process}.   

\begin{figure}[!tbp]
  \centering
  \begin{minipage}[b]{0.32\textwidth}
    \includegraphics[width=\textwidth]{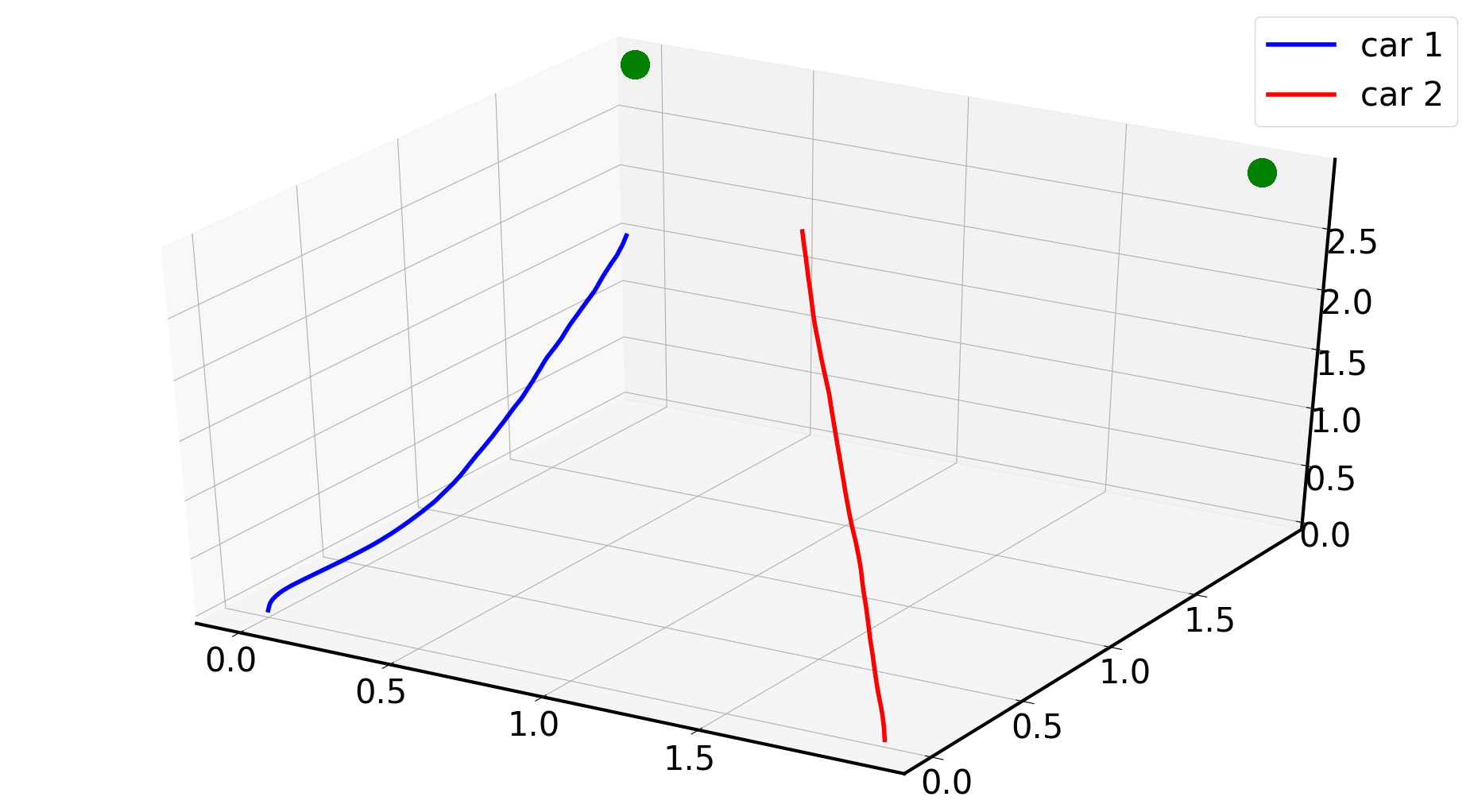}
  \end{minipage}
  \begin{minipage}[b]{0.32\textwidth}
    \includegraphics[width=\textwidth]{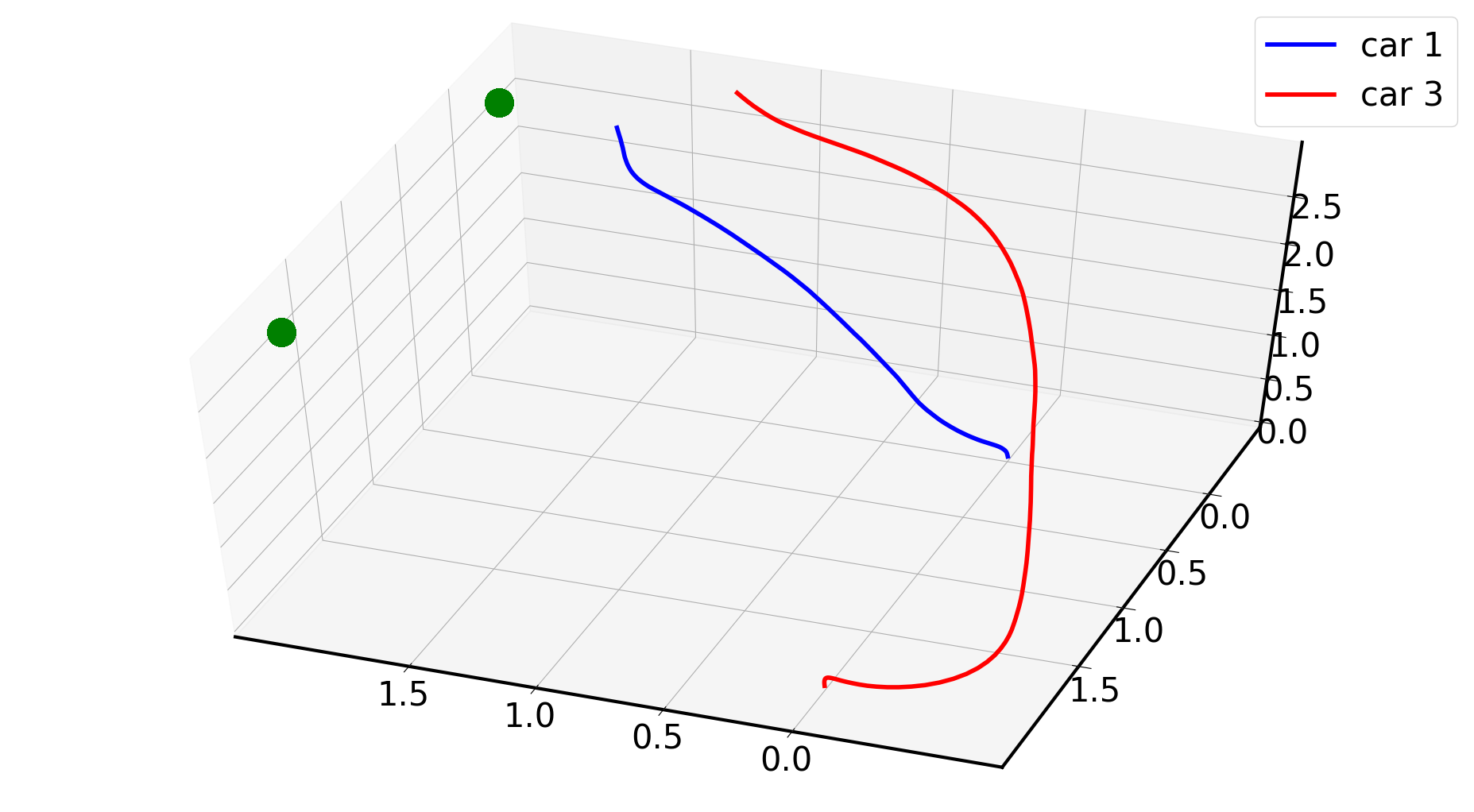}
  \end{minipage}
  \begin{minipage}[b]{0.32\textwidth}
    \includegraphics[width=\textwidth]{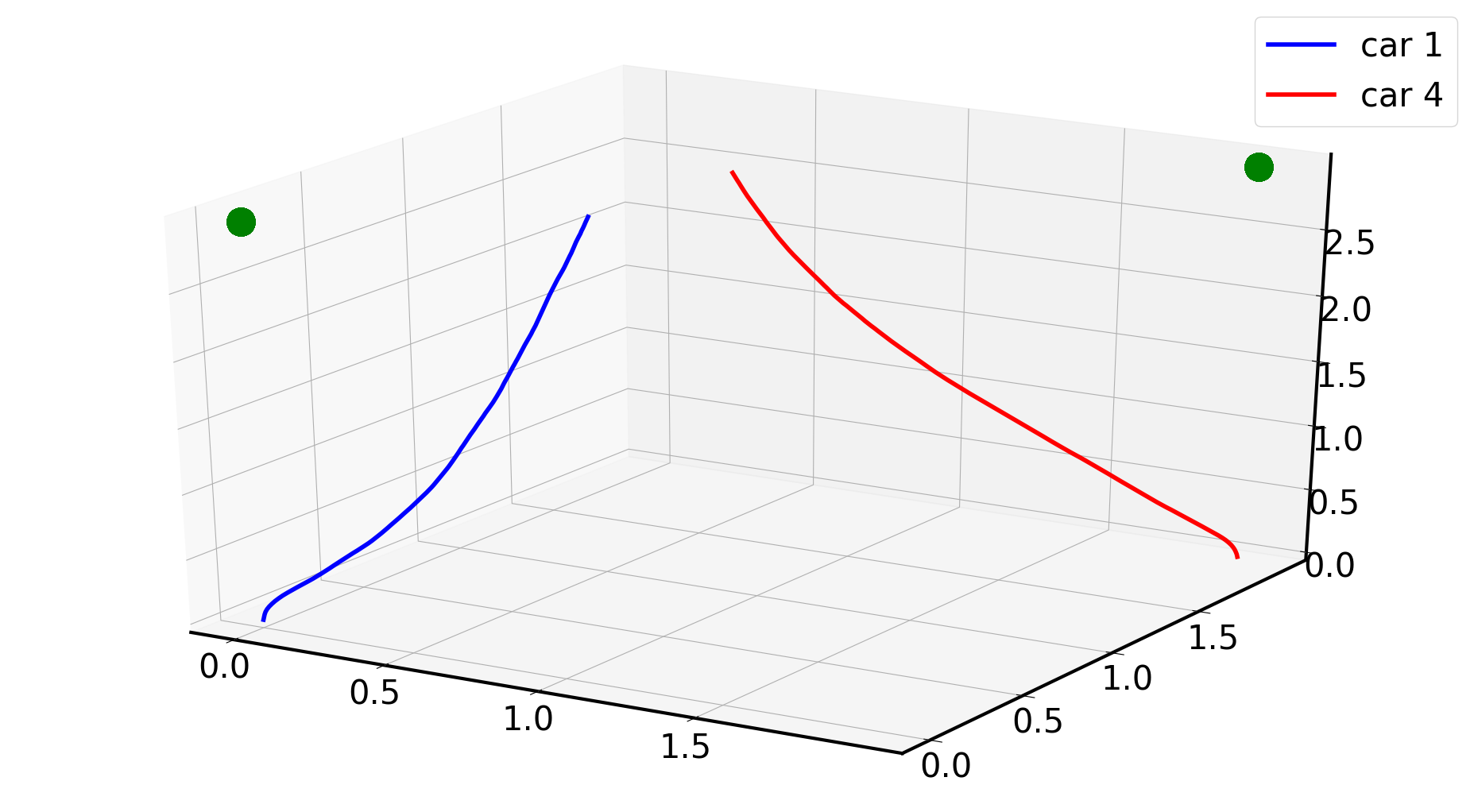}
  \end{minipage}
\end{figure}

\begin{figure}[!tbp]
  \centering
  \begin{minipage}[b]{0.32\textwidth}
    \includegraphics[width=\textwidth]{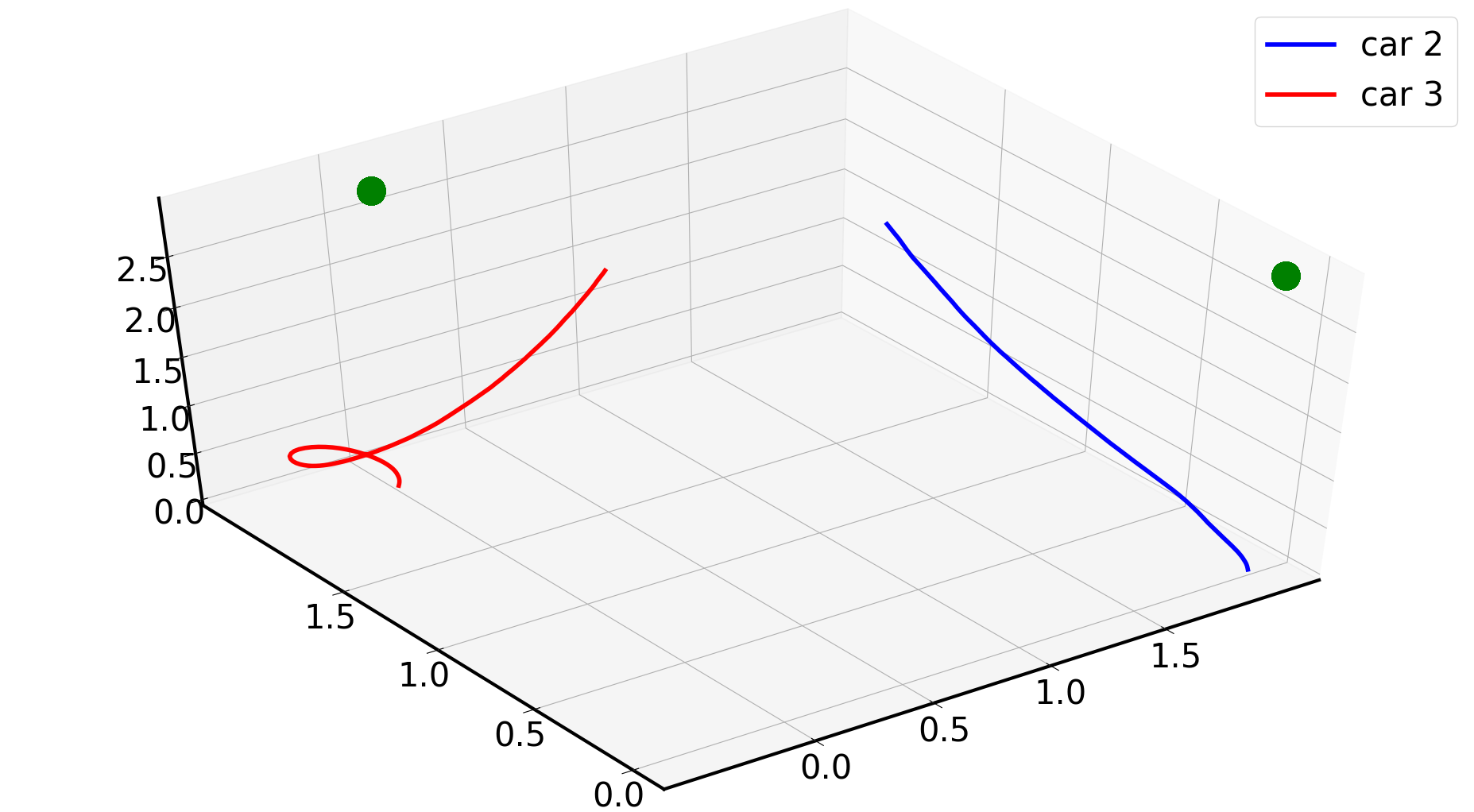}
  \end{minipage}
  \begin{minipage}[b]{0.32\textwidth}
    \includegraphics[width=\textwidth]{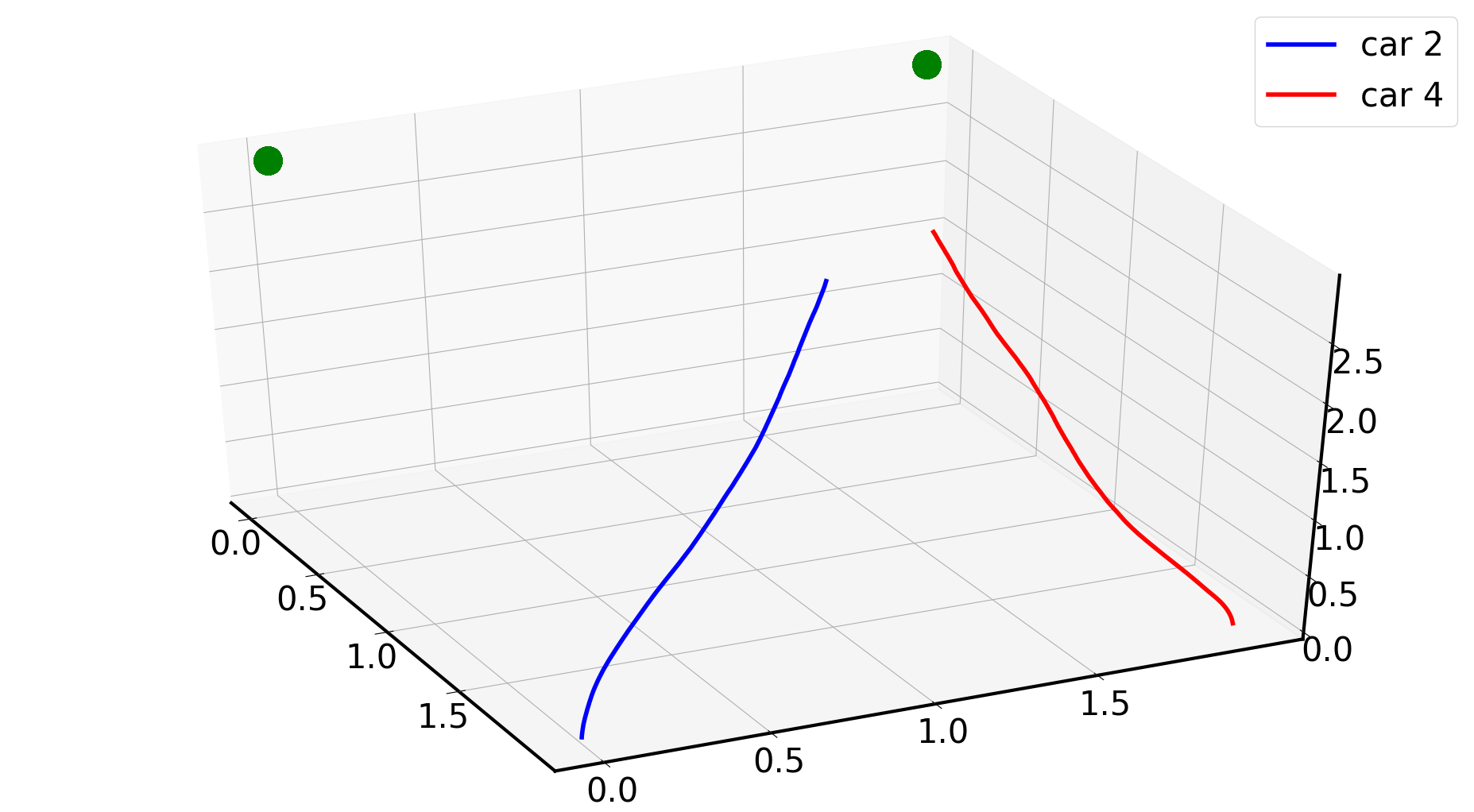}
  \end{minipage}
  \begin{minipage}[b]{0.32\textwidth}
    \includegraphics[width=\textwidth]{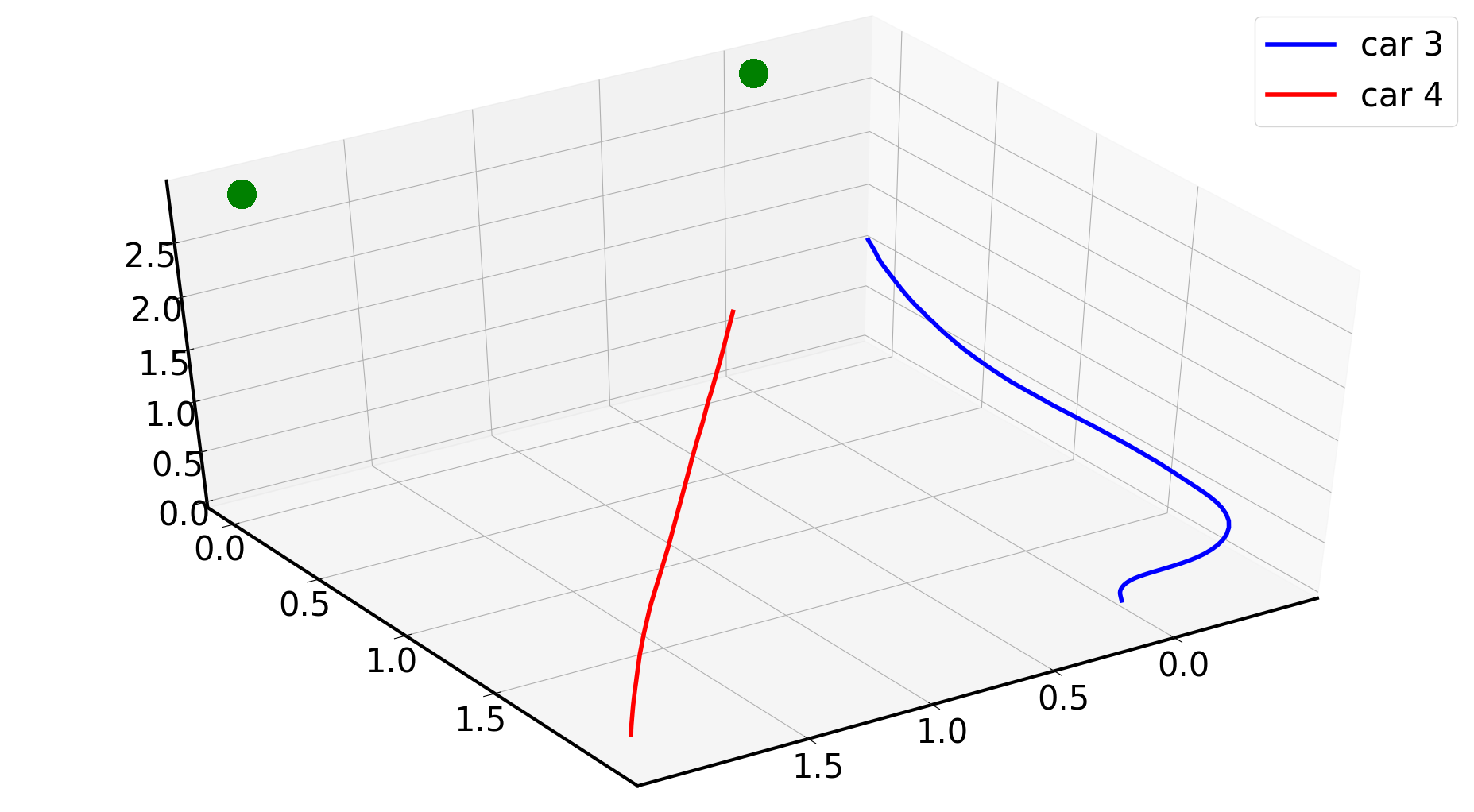}
  \end{minipage}
  \caption{\textbf{2D Car Collision Avoidance:} Figures above show worst-case (i.e. closest encounter) between each pair of cars during training. By design of the barrier function, each pair can be at 2 car radii from each other with zero velocity. This is seen above as the cars come to a halt much before reaching their targets. Vertical axes represent \textit{time (s)} and horizontal axes represent \textit{(x,y) position}.}
  \label{fig:2dcar_collision}
\end{figure}
\subsubsection{Pole-balancing}
Here we explored two sub-tasks (see Fig. \ref{fig:cartpole_balancing} for simulation results):
\par \textbf{With bounds on pole-angle}: Similar to the pendulum, the pole has to be stabilized at the unstable equilibrium point of $\theta=\pi$ rads. The initial state is $\vx=[0,\pi,0,0]^{\text{T}}$. Due to stochasticity, the pole is displaced off of $\theta=\pi$ and falls toward the stable equilibrium point of $\theta=0$. In this case (referred to as \textit{one constraint} case in Fig. \ref{fig:cartpole_balancing}(b) and Fig. \ref{fig:cartpole_balancing}(e)), the safe controller achieves the desired average performance during testing (\textit{red}) while never leaving the safe set of $\theta \in [\pi/2,3\pi/2]$ during training. Since the bounds on cart-position were not enforced, trajectories go beyond the limits as expected in Fig. \ref{fig:cartpole_balancing}(b).

\par \textbf{With bounds on both cart-position and pole-angle}:  The task here (referred to as \textit{double constraint} case in Fig. \ref{fig:cartpole_balancing}(c) and Fig. \ref{fig:cartpole_balancing}(f)) is identical to the one above, except now in addition to bounds on $\theta$, the cart-position is constrained to $x_c \in [-10\text{ m},\,10\text{ m}]$. Compared to the single constraint case, the safe controller now respects both bounds during training. 

Notice the similarity in Fig. \ref{fig:cartpole_balancing}(a) and Fig. \ref{fig:cartpole_balancing}(c), while Fig. \ref{fig:cartpole_balancing}(b) vastly differs. This is because, in the \textit{unsafe} Fig. \ref{fig:cartpole_balancing}(a), the optimal controller is allowed to let the pole fall beyond the safe bounds which does not encourage it to move the cart further away from the origin \textit{"to save the pole from falling"} like in Fig. \ref{fig:cartpole_balancing}(b). In Fig. \ref{fig:cartpole_balancing}(c), because of additional restrictions, the controller is forced to rapidly move the cart back-and-forth within the allowable bounds to stay safe. This becomes evident on comparing Fig. \ref{fig:cartpole_balancing}(e) and Fig. \ref{fig:cartpole_balancing}(f) wherein the range of $\theta$ covered is further reduced in the \textit{double constraint} case. Through this example we would like to highlight the adaptability of the safe FBSDE controller to find the best solution while respecting the imposed constraints.     

\subsection{Two-dimensional Car navigation}
\begin{wrapfigure}{r}{0.4\textwidth}
\vspace{-15 pt}
\centering
\includegraphics[width=0.4\textwidth]{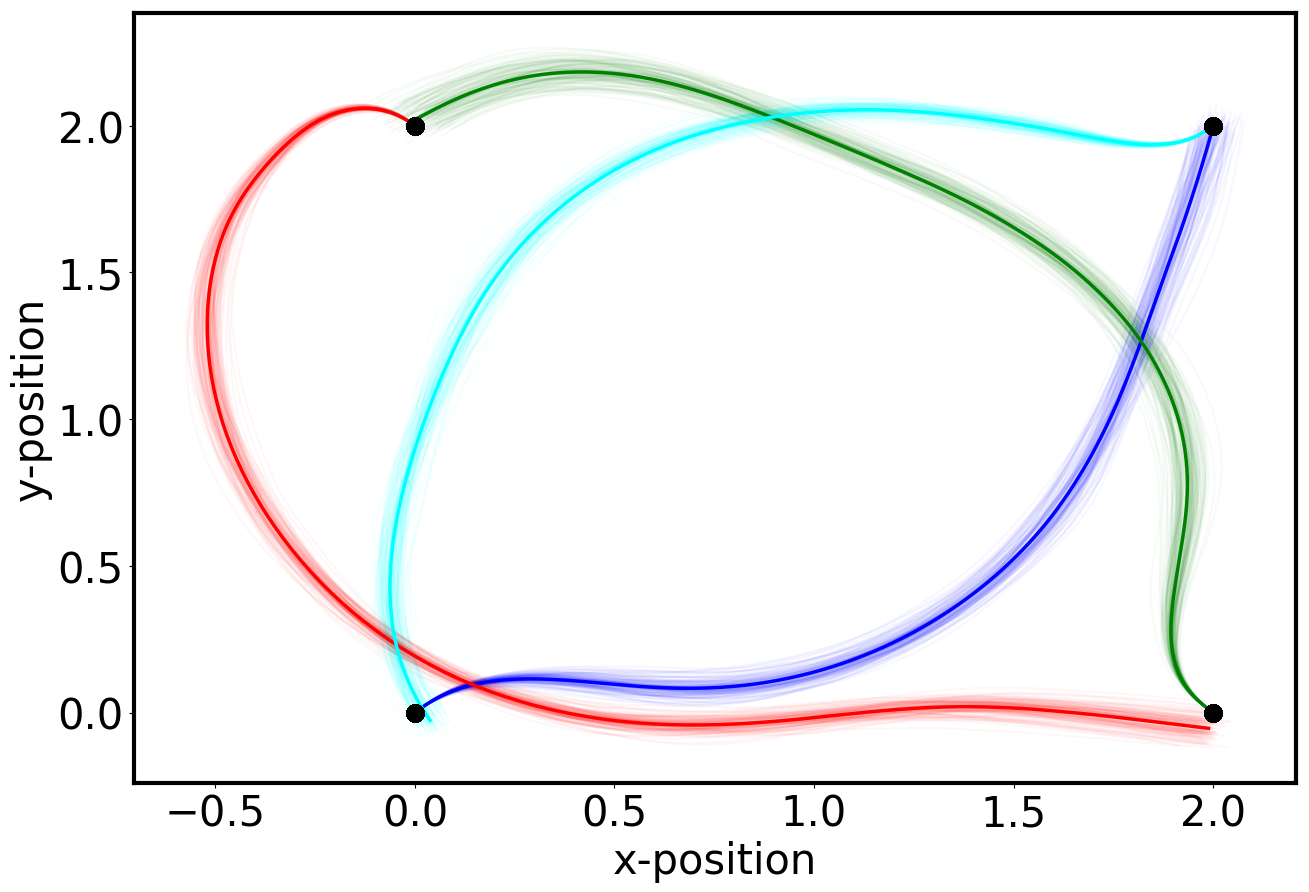}
  \caption{\textbf{Collision Avoidance:} Trajectories during testing learnt policy with darker lines indicating mean.}
\vspace{-20 pt}  
\label{fig:2dcar_collision_test}
\end{wrapfigure}
For each 2D car system, the state vector is given by $\vx=[p_x,p_y,\theta,v]^{\text{T}}$ representing the x-position, y-position, heading angle (with respect to the global x-axis) and the forward velocity of the car respectively. We consider the following two tasks:

\subsubsection{Single Car Multiple Obstacle Avoidance} In this task, a single car (see dynamics in Supplementary Material) has to navigate from an initial state of $\vx=[0,0,0,0]^{\text{T}}$ to a final state of $\vx=[2,2,0,0]^{\text{T}}$ while avoiding three obstacles shaped as circles whose centers are located at $(1,1),\,(1,0)$ and $(0,2)$ each with a radius $0.3$ in a time horizon of $3$ seconds. The simulation results are shown in Fig. \ref{fig:2dcar_obstacle}. Of the $256,256$ trajectories encountered during training, $10,000$ are randomly sampled and plotted as \textit{blue} lines. However, to emphasize that safety constraints \textit{are not violated during training}, we plot the worst-case trajectory (i.e. the trajectory that gets closest to any one of the three obstacles during training) in \textit{lime}. Notice that the optimal policy does not choose to be very close to the obstacle. This is due to the barrier function design, which allows the car to be at the obstacle boundary only if its velocity is zero. Therefore, getting closer to the boundary, slows the car down causing it to either not reach the target in the given time horizon or reach the target \textit{late} thereby accumulating more running cost. We hypothesize that during the training process, the learnt optimal cost-to-go (value function) of states very close to the obstacle are high and therefore the optimal controller chooses to stay away from the states close to the obstacle boundary. We hypothesize this also being the reason for the optimal policy choosing to go through the larger gap in between the upper and middle obstacle than the lower gap between the middle and lower obstacle.     

\subsubsection{Multi-Car Collision Avoidance} We finally tested our framework in a dynamic obstacle setting. In this task, 4 cars start at 4 corners of a square grid ($2\text{ m}\times2\text{ m}$) and are tasked with swapping their positions with the diagonally opposite car. To ensure safety, each car can be at a minimum distance of $0.1$ m from each other. Thus, the goal is to swap positions by avoiding collisions. We used ${4 \choose 2}=6$ barrier functions for each car pair.  Since all training trajectories cannot be plotted, we find the worst-cases (i.e. closest approach) between all car pairs during training. As seen in Fig. \ref{fig:2dcar_collision}, the cars stop short of their respective targets (\textit{green} spheres) because they get too close to each other. This behavior stems from our barrier function design, which allows the cars to be at $0.1$ m from each other but with zero velocity. We hypothesize that these states get associated with high value (high cost-to-go) which the optimal controller learns to avoid. The performance during testing is shown in Fig. \ref{fig:2dcar_collision_test} wherein each car successfully reaches its target on average. Please refer to the supplementary for 3D plots during testing to see successful collision avoidance.  An animation \footnote{https://youtu.be/noonPzL39ic} demonstrating the progress of the policy through the training process is available on YouTube.

\section{Conclusions and Future Directions}
\label{future_work}
In this paper we introduced a novel approach to safe reinforcement learning for stochastic dynamical systems. Our framework is the first to integrate Stochastic CBFs into the deep FBSDE controller framework for end-to-end safe optimal control. We extended the existing theory on Stochastic ZCBFs, developed a numerical algorithm combining deep FBSDEs, stochastic ZCBFs and the differentiable convex optimization layer (OptNet) frameworks and successfully demonstrated its application on three systems in simulation. This initial success motivates a few directions for further investigation. Namely, we would like to explore high relative degree stochastic CBFs for highly underactuated systems and combine this framework for systems in simulators for which explicit equations of motion are not available. We believe these would serve as stepping stones to be able to deploy this algorithm on real hardware in the future. 

\section{Acknowledgements}
This work is supported by NASA and the NSF-CPS \#1932288

\bibliography{references}  

\newpage
\section{Supplementary Material}
\subsection{Proof of Theorem 1}
Theorem 1 is a generalization of Theorem [3] in \cite{clark2020control}, the proof of which is mimicked in construction here. We want to show that for any $t>0$, any $\epsilon>0$, and any $\delta \in (0,1)$,
$$\mathbb{P}\big(\underset{t'<t}{\inf}h(\vx_{t'})\leq-\epsilon)\leq\delta, $$
where the subscript $t$ denotes dependence on time. Let $\theta = \min \{\alpha^{-1}\big(\frac{\delta\epsilon}{2t}\big),h(\vx_0)\}$. Note that for $t=0$ this reduces to 
\begin{equation}\label{h0}
h(\vx_0)=\theta.
\end{equation}
It\^o's lemma then implies that
\begin{equation}\label{ITOH}
    h(\vx_t)=h(\vx_0)+\int_0^t\frac{\partial h}{\partial \vx}(\vf(\vx_{\tau}) +\vG(\vx_{\tau})\vu)+\frac{1}{2}\tr\big(\frac{\partial^2h}{\partial\vx^2}\Sigma\Sigma^{\text{T}}\big)\mathrm{d}\tau+\int_0^t\Sigma(\vx_{\tau})\frac{\partial h}{\partial \vx }\mathrm{d}\vw_{\tau}.
\end{equation}
Consider now $\theta$ as a particular value for $h(\vx)$. As $h(\vx)$ varies, it may cross this value from above or from bellow. We call the times in which this occurs \textit{stopping times}; in particular, we have stopping times $\eta_i$ whenever $h(\vx_t)$ down-crosses the value $\theta$ (i.e., coming from above) and $\zeta_i$ whenever $h(\vx_t)$ up-crosses the value $\theta$ (i.e., coming from below). The sequence is defined as follows:
\begin{align*}
    &\eta_0 = 0, &\zeta_0=\inf\{t:h(\vx_t)>\theta\},\\
    &\eta_i = \inf\{t:h(\vx_t)<\theta,~t>\zeta_{i-1}\}, &\zeta_i = \inf\{t:h(\vx_t)>\theta,~t>\eta_{i-1}\},
\end{align*}
for $i=1,2,\hdots$. Note that since $h(\vx_0)=\theta$, we take $\eta_0=0$ by convention. We now proceed to construct a random process $U_t$ with $U_0=\theta$ as follows:
\begin{equation}\label{Ut}
    U_t = \theta + \sum_{i=0}^{\infty}\bigg( \int_{\eta_i\wedge t}^{\zeta_i\wedge t}-\alpha(\theta)\mathrm{d}\tau+\int_{\eta_i\wedge t}^{\zeta_i\wedge t}\Sigma\frac{\partial h}{\partial \vx}\mathrm{d}\vw_{\tau}\bigg),
\end{equation}
in which $a\wedge b$ denotes the minimum between $a$ and $b$. For $s<t$, this process satisfies
\begin{align*}
    \mathbb{E}[U_t|U_s] &= U_s + \mathbb{E}\bigg[\sum_{i=0}^{\infty}\bigg( \int_{(\eta_i\wedge t)\vee s}^{(\zeta_i\wedge t)\vee s}-\alpha(\theta)\mathrm{d}\tau+\int_{(\eta_i\wedge t)\vee s}^{(\zeta_i\wedge t)\vee s}\Sigma\frac{\partial h}{\partial \vx}\mathrm{d}\vw_{\tau}\bigg) \bigg]\\
    &=U_s +  \mathbb{E}\bigg[\sum_{i=0}^{\infty}\int_{(\eta_i\wedge t)\vee s}^{(\zeta_i\wedge t)\vee s}-\alpha(\theta)\mathrm{d}\tau\bigg] \leq U_s,
\end{align*}
where $a\vee b$ denotes the maximum between $a$ and $b$, and  $(\eta_i\wedge t)\vee s$ is used to avoid integration during times up to $s$, which are already included in $U_s$. This inequality implies that the stochastic process $U_t$ is a \textit{supermartingale}.

We will now show that (a). $h(\vx_t)\geq U_t$ and (b). $U_t\leq \theta$ for all $t$. This will be done by induction. For $t=\eta_0=0$, both statements hold, because $h(\vx_0)=\theta$ as per \eqref{h0}, and $U_0=\theta$ by construction (eq. \eqref{Ut}). Now, suppose that the statements hold up to time $t=\eta_i$ for some $i\geq0$. We will show that they remain true for the interval $[\eta_i,\zeta_i]$ first, and then that they remain true for the interval $[\zeta_{i},\eta_{i+1}]$, which would complete the induction. For $t\in [\eta_i,\zeta_i]$ we have
\begin{align*}
    &h(\vx_t) =h(\vx_{\eta_i})+\int_{\eta_i}^t\frac{\partial h}{\partial \vx}(\vf(\vx_{\tau}) +\vG(\vx_{\tau})\vu)+\frac{1}{2}\tr\big(\frac{\partial^2h}{\partial\vx^2}\Sigma\Sigma^{\text{T}}\big)\mathrm{d}\tau+\int_{\eta_i}^t\Sigma(\vx_{\tau})\frac{\partial h}{\partial \vx }\mathrm{d}\vw_{\tau},\\
    &U_t = U_{\eta_i}+\int_{\eta_i}^{t}-\alpha(\theta)\mathrm{d}\tau+\int_{\eta_i}^{t}\Sigma(\vx_{\tau})\frac{\partial h}{\partial \vx }\mathrm{d}\vw_{\tau}.
\end{align*}
We perform comparison by terms: for the first terms, $h(\vx_{\eta_i})\geq U_{\eta_i}$ by the main assumption of our induction process. The third terms are common, and the integrands of the second terms relate as follows:
\begin{equation*}
    \frac{\partial h}{\partial \vx}(\vf(\vx) +\vG(\vx)\vu)+\frac{1}{2}\tr\big(\frac{\partial^2h}{\partial\vx^2}\Sigma\Sigma^{\text{T}}\big) \geq -\alpha(h(\vx))\geq-\alpha(\theta),
\end{equation*}
where the first inequality is due to the CBF inequality \eqref{CBFineq} and the second inequality is by definition of stopping times $\eta_i$ and $\zeta_i$ and monotonicity of $\alpha(\cdot)$: in the interval $[\eta_i,\zeta_i]$ we have that $h(\vx_t)\leq\theta$. Thus, for the interval $[\eta_i,\zeta_i]$, we showed that  $h(\vx_t)\geq U_t$, and as a corollary, by virtue of the stopping time definition, we furthermore have $U_t\leq \theta$. We proceed with the interval $t\in [\zeta_{i},\eta_{i+1}]$. By construction, $U_t$ remains constant during that interval and we have
$$U_t =U_{\zeta_i}\leq h(x_{\zeta_i})=\theta\leq h(\vx_t),$$
wherein the first inequality holds as a result of the previous interval ($t\in[\eta_i,\zeta_i]$), and the second equality and inequality hold by virtue of the definition of the stopping times $\zeta_i$ and $\eta_{i+1}$. This completes the induction and we have thus proved that for all $t$, we have $h(\vx_t)\geq U_t$ and $U_t\leq \theta$.

Since $U_t\leq h(\vx_t)$, it follows that
$$\mathbb{P}\big(\underset{t'<t}{\inf}h(\vx_{t'})\leq-\epsilon)\leq\mathbb{P}\big(\underset{t'<t}{\inf}U_{t'}\leq-\epsilon\big). $$ 
By Doob's supermartingale inequality \cite{karatzas1998brownian} we have that for $\epsilon>0$ 
$$\epsilon \mathbb{P}\big(\underset{t'<t}{\inf}U_{t'}\leq-\epsilon\big)\leq \mathbb{E}[U_t^+]-\mathbb{E}[U_t],$$
wherein $a^+=\max\{a,0\}$. Taking $s=\max\{i:\eta_i<t\}$, we can express $\mathbb{E}[U_t]$ as
\begin{equation*}
\mathbb{E}[U_t]=
\left\{
\begin{array}{ll}
      \theta-\alpha(\theta)\mathbb{E}[\sum_{i=0}^s(\zeta_i-\eta_i)], & t\in[\zeta_s,\eta_{s+1}], \\
      \theta-\alpha(\theta)\mathbb{E}[t-\eta_s+\sum_{i=0}^{s-1}(\zeta_i-\eta_i)], & t\in[\eta_s,\zeta_{s}]. \\
      \end{array} 
\right. 
\end{equation*}
Both expectations inside the bracket are bounded above by $t$, and thus $\mathbb{E}[U_t]\geq\theta-\alpha(\theta)t$. Furthermore, since $U_t\leq \theta$ we have $\mathbb{E}[U_t^+]\leq\theta$, so combining the two we obtain $$\mathbb{E}[U^+_t]-\mathbb{E}[U_t]\leq\theta-(\theta-\alpha(\theta)t)=\alpha(\theta)t.$$
Thus,
\begin{align*}
    \mathbb{P}\big(\underset{t'<t}{\inf}h(\vx_{t'})\leq-\epsilon)&\leq\mathbb{P}\big(\underset{t'<t}{\inf}U_{t'}\leq-\epsilon\big)\\&\leq\alpha(\theta)\frac{t}{\epsilon}\leq\alpha\big(\alpha^{-1}(\frac{\delta \epsilon}{2t})\big)\frac{t}{\epsilon}=\frac{\delta \epsilon t}{2\epsilon t}< \delta,
\end{align*} 
where we used that $\theta =\min \{\alpha^{-1}\big(\frac{\delta\epsilon}{2t}\big),h(\vx_0)\}$ is no greater than $\alpha^{-1}\big(\frac{\delta\epsilon}{2t}\big)$. The proof is concluded.

\subsection{Relative Degree of Barrier Functions}
Let $L_fh$ and $L_g h$ be the Lie derivatives of $h$, defined as $\frac{\partial h}{\partial \vx}\vf$ and $\frac{\partial h}{\partial \vx}\vG$, respectively. The relative degree $r$ of a barrier function is defined such that $L_g L_f^{r-1}h(\vx)\neq 0$ and $L_g h(\vx)=L_g L_f h(\vx)=L_g L_f^2 h(\vx)=\ldots=L_g L_f^{r-2} h(\vx)=0, \quad \forall \vx$. Note that relative degree 1 barrier functions are  barrier functions that impose restrictions on directly actuated system states, as a result of which $L_gh(\vx)\neq 0$. 

\subsection{Implementation Details}
In all our simulations we pick the following form of the class $\mathcal{K}$ function that bounds the rate of change of the barrier function, $\alpha(\cdot) = \gamma h(\vx)$, where $\gamma$ is a fixed positive constant that changes the slope of the bounding function. Higher the value of $\gamma$, less conservative is the barrier function.  
\subsubsection{Inverted Pendulum}
For the inverted pendulum system, the state vector is given by $\vx=[\theta, \dot{\theta}]^{\text{T}}$ indicating pendulum angular-position and pendulum angular-velocity. The system dynamics $(f)$ vector and the actuation $(G)$ matrix are given by, 
$$f=\begin{bmatrix} \dot{\theta} \\ \dfrac{-b}{I}\dot{\theta} - \dfrac{g}{l}\sin{\theta} \end{bmatrix}, \quad G = \bigg[0, \dfrac{1}{I}\bigg]^{\text{T}} $$
\par The stochastic dynamics are given by, $$\text{d}\vx(t) = f(\vx)\,\text{d}t + G(\vx)u\, \text{d}t + \Sigma(\vx)\, \text{d}W(t)$$ where, $$\Sigma = \begin{bmatrix} 0 & 0 \\ 0 & \sigma \end{bmatrix}\implies \Sigma\Sigma^{\text{T}}=\begin{bmatrix} 0 & 0 \\ 0 & \sigma^2  \end{bmatrix}$$
The parameter values used were: $b=0.1$, $l=0.5$ m, $g=9.81\,\text{m/s}^2$, $m=2$ kg and $I=ml^2$.
\par \textbf{Balancing with angle constraints:} For this task, we impose box constraints on the pendulum angular position. The goal is to ensure that the pendulum does not fall outside of a predetermined safe region around the unstable equilibrium point of $\vx=[\pi,0]^{\text{T}}$. Although the pendulum is initialized at $\vx=[\pi,0]^{\text{T}}$, stochasticity in the dynamics will push it off this unstable equilibrium position and safe controller is then tasked to keep the pendulum inside the safe region. As stated earlier, the pendulum must remain inside the safe region \textit{during the entire training process}. 
\par Let $\theta_h$ and $\theta_l$ be the upper and lower bounds on the angular position. Let the corresponding barrier functions be $h_h(\vx) = \theta_h - \theta$ and $h_l(\vx) = \theta - \theta_l$ and the combined barrier function be $h(\vx) = h_h(\vx)\cdot h_l(\vx)-\mu\dot{\theta}^2$, so that the safe set given by $\mathcal{C}=\{\vx:h(\vx)\geq 0\}$.
    
\begin{align*}
    h(\vx) &= (\theta_h - \theta)\cdot(\theta - \theta_l)-\mu\dot{\theta}^2 = \theta_h \theta - \theta_h \theta_l - \theta^2 + \theta_l \theta -\mu\dot{\theta}^2\\
    \dfrac{\partial h}{\partial \vx} &=\begin{bmatrix} \theta_h - 2 \theta + \theta_l \\ -2\mu\dot{\theta}\end{bmatrix} \quad  \implies \dfrac{\partial h}{\partial \vx}^{\text{T}} G = \dfrac{-2\mu\dot{\theta}}{I}\\
    \therefore C &= -\dfrac{\partial h}{\partial \vx}^{\text{T}} G=\dfrac{2\mu\dot{\theta}}{I} \quad \cdots\text{ for } \eqref{eq:qp}\\
    \dfrac{\partial h}{\partial \vx}^{\text{T}}f &= \dot{\theta}(\theta_h - 2 \theta + \theta_l) + \bigg(\dfrac{-b}{I}\dot{\theta} - \dfrac{g}{l}\sin{\theta}\bigg) \cdot (-2\mu\dot{\theta})\\
    &= \dot{\theta}(\theta_h - 2 \theta + \theta_l) + \bigg(\dfrac{b}{I}\dot{\theta} + \dfrac{g}{l}\sin{\theta}\bigg) \cdot (2\mu\dot{\theta})\\
    \therefore d &= \alpha(h(\vx)) + \dfrac{1}{2}\text{tr}\bigg(\dfrac{\partial^2 h}{\partial \dot{\theta}^2}\Sigma\Sigma^{\text{T}}\bigg) + \dfrac{\partial h}{\partial \vx}^{\text{T}}f \quad \cdots\text{ for } \eqref{eq:qp}\\
    &= \alpha(h(\vx)) - \mu \sigma^2  + \dfrac{\partial h}{\partial \vx}^{\text{T}}f \quad \cdots \bigg( \because \dfrac{\partial^2 h}{\partial \dot{\theta}^2} = -2\mu\bigg) 
\end{align*}
\par The above choice of barrier function is relative degree 1 because $L_gh = (\partial h/\partial \vx)^\text{T}g \neq 0$, except at $\dot{\theta}=0$. However, given the task at hand, the only time this can occur is at initial condition $\vx=[\pi,0]^\text{T}$ or if the controller manages to stabilize the pendulum at the unstable equilibrium point and exactly cancel out all the stochasticity entering the system. If any one of these occur, since $\theta=\pi$ (and because we choose box-constraint like bounds eg. $(\theta_l,\theta_h)=(2\pi/3,4\pi/3)$), we see that the gradient of the barrier function $\partial h/\partial \vx$ is itself zero. This means that the safety constraint is trivially satisfied (as we have $0 \geq -\alpha(h(\vx))$ in \eqref{CBFineq}. In these cases, the unconstrained optimal control, $\vu^*(t,\vx)=-\vR^{-1}\vG\T V_{\vx}$, would be the solution to the QP problem.
\par Regarding values of hyper-parameters used to train the Safe FBSDE controller for this problem, we used the following,
\begin{center}
 \begin{tabular}{||c| c| c||} 
 \hline
 \# & \textbf{Parameter Name} & \textbf{Parameter Value} \\  
 \hline\hline
 1 & batch size & 128 \\ 
 \hline
 2 & training iterations & 201 \\
 \hline
 3 & $\sigma$ & 1.0 \\
 \hline
 4 & learning rate & $1e^{-2}$ \\
 \hline
 5 & number of neurons per LSTM layer & 16 \\
 \hline
 6 & optimizer & Adam \\
 \hline
 7 & $\mu$ & $0.05$ \\
 \hline
 8 & $\gamma$ & $0.5$\\
 \hline
\end{tabular}
\end{center}

\subsubsection{Cart Pole}
For this system, the state vector $\vx=\big[x_c,\, \theta,\, \dot{x_c},\, \dot{\theta}\big]^{\text{T}}$, indicating cart-position, pole angular-position, cart-velocity and pole angular-velocity respectively. The system dynamics ($f$) vector and actuation ($G$) matrix are given by, $$f=\begin{bmatrix} \dot{x_c} \\ \dot{\theta} \\ \dfrac{m_p \sin{\theta}(l \dot{\theta}^2 + g \cos{\theta})}{m_c + m_p \sin^2{\theta}} \\ \dfrac{-m_p l  \dot{\theta}^2 \cos{\theta} \sin{\theta} - (m_c+m_p) g \sin{\theta}}{l(m_c + m_p \sin^2{\theta})} \end{bmatrix} = \begin{bmatrix} f_1 \\ f_2 \\ f_3 \\ f_4 \end{bmatrix}, \quad G=\begin{bmatrix} 0 \\ 0 \\ \dfrac{1}{m_c + m_p \sin^2{\theta}} \\ \dfrac{-\cos{\theta}}{l(m_c + m_p \sin^2{\theta})}\end{bmatrix} = \begin{bmatrix} 0 \\ 0 \\ g_3 \\ g_4 \end{bmatrix}$$
The stochastic dynamics are given by, $$\text{d}\vx(t) = f(\vx)\,\text{d}t + G(\vx)u\, \text{d}t + \Sigma(\vx)\, \text{d}W(t)$$ where, $$\Sigma(\vx) = \begin{bmatrix} 0 & 0 & 0 & 0 \\ 0 & 0 & 0 & 0 \\ 0 & 0 & \sigma & 0 \\ 0 & 0 & 0 & \sigma \end{bmatrix} \implies \Sigma(\vx) \Sigma^{\text{T}}(\vx)=\begin{bmatrix} 0 & 0 & 0 & 0 \\ 0 & 0 & 0 & 0 \\ 0 & 0 & \sigma^2 & 0 \\ 0 & 0 & 0 & \sigma^2 \end{bmatrix}$$
The system parameters chosen were: $m_p=0.01$ kg, $m_c=1.0$ kg, $g=9.81\,\text{m/s}^2$ and $l=0.5$ m.
\begin{enumerate}
    \item \textbf{Swing-up with cart position constraints:} Here the task is to begin from the initial position $\vx_0=[0,0,0,0]^{\text{T}}$ and perform a swing-up to $\vx_{\text{target}} = [0,\pi,0,0]^{\text{T}} $ within a fixed time horizon of $1.5$ seconds. Additionally, there are constraints on the cart-position that need to be satisfied both during training (i.e. learning the policy) and testing (i.e. final deployment of the policy). These constraints are represented as $[x_{c,l},\, x_{c,h}]$ where $x_{c,h}>x_{c,l}$. 
    A choice of barrier function that combines both position and velocity constraints is as follows: $$\tilde{h}=(x_{c,h}-x_c)\cdot(x_c  -x_{c,l}) - \mu \dot{x_c}^2 = x_{c,h}\,x_c - x_{c,h}\,x_{c,l} - x_c^2 + x_c\, x_{c,l} - \mu \dot{x_c}^2$$ where the parameter $\mu$ controls \textit{how fast} the cart is allowed to move in the interior of the safe set $\mathcal{C}$. 
    \begin{align*}
        \dfrac{\partial \tilde{h}}{\partial \vx} &= \begin{bmatrix} x_{c,h} - 2x_c + x_{c,l} \\ 0 \\ -2\mu \dot{x_c} \\ 0  \end{bmatrix} \implies \dfrac{\partial \tilde{h}}{\partial \vx}^{\text{T}}G = \dfrac{-2\mu\dot{x_c}}{m_c + m_p \sin^2{\theta}} \\
        \therefore \, C &= -\dfrac{\partial \tilde{h}}{\partial \vx}^{\text{T}}G \quad \text{ and }\\
        d &= \alpha(\tilde{h}(\vx)) + 0.5\cdot\sigma^2(-2\mu) + \dfrac{\partial \tilde{h}}{\partial \vx}^{\text{T}}f \quad\cdots \bigg(\dfrac{\partial^2 \tilde{h}}{\partial \dot{x_c}^2}=-2\mu\bigg)
    \end{align*}
    The relative degree of this barrier is also 1 similar to the pendulum case above. Please see the explanation provided there. For this task we used, $\mu=0.1$, $\gamma=1.0$ and cart-position bounds of $(x_{c,l},x_{c,h})=(-10\text{ m},10\text{ m})$.
    \item \textbf{Pole balancing with pole position constraints:}
    Similar to the construction above for position constrained swing-up, we have the following hybrid barrier function for the task of balancing with constrained pole angle, $$h_\theta(\vx) = (\theta_h - \theta)\cdot(\theta - \theta_l)- \mu_{\theta}\dot{\theta}^2=\theta_h\theta -\theta_h \theta_l - \theta^2 + \theta \theta_l - \mu_{\theta}\dot{\theta}^2$$
    \begin{align*}
        \dfrac{\partial h_\theta}{\partial \vx} &= \begin{bmatrix} 0 \\ \theta_h - 2\theta + \theta_l \\ 0 \\ -2\mu_\theta\dot{\theta} \end{bmatrix} \implies \dfrac{\partial h_\theta}{\partial \vx}^{\text{T}}G = \dfrac{(-2 \mu_\theta \dot{\theta})\cdot(-\cos{\theta})}{l(m_c+m_p\sin^2{\theta})}\\
        \therefore \, C &= -\dfrac{\partial h_\theta}{\partial \vx}^{\text{T}}G \quad \text{ and }\\
        d &= \alpha(h_\theta(\vx)) + 0.5\cdot\sigma^2(-2\mu_\theta) + \dfrac{\partial h_\theta}{\partial \vx}^{\text{T}}f \quad\cdots \bigg(\dfrac{\partial^2 h}{\partial \dot{\theta}^2}=-2\mu_\theta\bigg)
    \end{align*}
    The relative degree of this barrier is also 1 similar to the pendulum case above. Please see the explanation provided there. For this task we used, $\mu=0.1$, $\gamma=1.0$ and pole angular-position bounds of $(\theta_l,\theta_h)=(\pi/2\text{ rad},3\pi/2\text{ rad})$.
    \item \textbf{Pole balancing with both pole and cart position constraints:} For this task we consider two separate barrier functions i.e. two safety inequality constraints for cart-position and pole-angle. The barrier functions and corresponding terms of the inequality constraints are exactly the same as the two sub-tasks above. The bounds are also the same as those used in the above sub-tasks. The barrier function parameters used were $\mu_{x_c}=0.01$, $\mu_\theta=10$, $\gamma_{x_c}=10$ and $\gamma_\theta=100$.
\end{enumerate}
\par Following are some of the common hyper-parameter values used for the above cart-pole simulations:
\begin{center}
 \begin{tabular}{||c| c| c||} 
 \hline
 \# & \textbf{Parameter Name} & \textbf{Parameter Value} \\  
 \hline\hline
 1 & batch size & 128 \\ 
 \hline
 2 & training iterations (tasks i.) & 2001 \\
 \hline
 3 & training iterations (tasks ii. and ii.) & 201 \\
 \hline
 4 & $\sigma$ & 1.0 \\
 \hline
 4 & learning rate & $1e^{-2}$ \\
 \hline
 5 & number of neurons per LSTM layer & 16 \\
 \hline
 6 & optimizer & Adam \\
 \hline
\end{tabular}
\end{center}

\subsubsection{2-D Car}
 The model is same as that used in \cite{xie2017} and has the following dynamics:
\begin{align*}
    \dot{x}=v \cos(\theta),\quad &
  \dot{y}=v \sin(\theta)\\
  \dot{\theta}= vu^{\theta}, \quad &
  \dot{v}= u^{v}
\end{align*}
The state vector is $\vx=[p_x,p_y, \theta, v]\T$ which corresponds to the x-position, y-position, heading angle and forward velocity of the car. The control vector is $\vu=[u^\theta, u^v]\T$ which corresponds to the acceleration and steering-rate of the car. The system dynamics $(f)$ vector and actuation $(G)$ matrix are given by,
\begin{equation*}
    f=\begin{bmatrix}
    v\cos(\theta)\\v\sin(\theta)\\0\\0
    \end{bmatrix},\quad G=\begin{bmatrix}
    0 & 0\\0 & 0\\ v & 0 \\ 0 & 1
    \end{bmatrix}
\end{equation*}
The stochastic dynamics are given by,
\begin{equation*}
    \rd \vx = f(\vx)\rd t + g(\vx)\vu\rd t + \Sigma(\vx) \rd W
\end{equation*}
where
\begin{equation*}
    \Sigma(\vx) = \begin{bmatrix} 0 & 0 & 0 & 0 \\ 0 & 0 & 0 & 0 \\ 0 & 0 & \sigma & 0 \\ 0 & 0 & 0 & \sigma \end{bmatrix} \implies \Sigma(\vx)\Sigma\T(\vx) = \begin{bmatrix} 0 & 0 & 0 & 0 \\ 0 & 0 & 0 & 0 \\ 0 & 0 & \sigma^2 & 0 \\ 0 & 0 & 0 & \sigma^2 \end{bmatrix}
\end{equation*}
\begin{enumerate}

    \item \textbf{Single Car Multiple Obstacle Avoidance:} In this task a single car starts at $(p_x, p_y)=(0\text{ m},0\text{ m})$ with the goal to reach the target $(p_x, p_y)=(2\text{ m},2\text{ m})$ while avoiding obstacles shaped as circles. We use 1 barrier function for each obstacle which takes the following form:
    \begin{equation*}
        \tilde{h}^i(\vx)=(p_x-o^i_x)^2+(p_y-o^i_y)^2-{o^i_r}^2-\mu v^2
    \end{equation*}
    where $i$ is the obstacle index and $o^i_x, o^i_y, o^i_r$ are the x-position, y-position and radius of the obstacle respectively. This is a relative degree 1 barrier function, which can be shown by
        \begin{align*}
        \frac{\partial \tilde{h}^i}{\partial \vx}&=\begin{bmatrix}
        2(p_x - o^i_x)\\
        2(p_y - o^i_y)\\ 0\\-2\mu v 
        \end{bmatrix}\\
        \implies \frac{\partial \tilde{h}^i}{\partial \vx}\T G &=  \begin{bmatrix}0 & -2\mu v \end{bmatrix}
    \end{align*}
    The other terms needed to set up \eqref{eq:qp} can be calculated as:
    \begin{align*}
        \frac{\partial \tilde{h}^i}{\partial \vx}\T f &= 2(p_x - o^i_x)v\cos(\theta) + 2(p_y - o^i_y)v\sin(\theta)\\
        \frac{1}{2}\tr\big( \frac{\partial^2 \tilde{h}^i}{\partial \vx^2} \Sigma\Sigma\T\big) &= -\mu\sigma^2
    \end{align*}
    The QP can be set up with
    \begin{align*}
        C_i &= -\dfrac{\partial \tilde{h}^i }{\partial \vx}^{\text{T}}G\\
        d_i &= \alpha(\tilde{h}^i(\vx))+\dfrac{1}{2}\tr \big( \dfrac{\partial^2 \Tilde{h}^i}{\partial \vx^2}\Sigma\Sigma\T\big) + \frac{\partial \tilde{h}^i}{\partial \vx}^{\text{T}}f
    \end{align*}
    \item \textbf{Multi-car Collision Avoidance:}
    For multi-car collision avoidance, simply set the obstacle's positions in the previous section's barrier function to the other car's positions, the obstacle's size to be 2 times the car's size, and subtract an additional velocity of the other car. In our simulation example we used 4 cars for a total of 16 states, $\vx=[p_{x_1}, p_{y_1},\dots,\theta_4, v_4]\T$, dimensions and 8 control dimensions, $\vu=[u^\theta_1, u^v_1,\dots,u^\theta_4,u^v_4]\T$, resulting in 6 barrier functions for each pair of vehicles. The initial conditions for the 4 cars are $[0,0,\pi/4,0.1]$, $[2,0,3\pi/4,0.1]$, $[0,2,-\pi/4,0.1]$ and $[2,2,-3\pi/4,0.1]$ respectively. For example, the barrier between car 1 and  car 2 looks like
    \begin{equation*}
        \tilde{h}^{12}(\vx)=(p_{x_1}-p_{x_2})^2+(p_{y_1}-p_{y_2})^2-(2o_c)^2-\mu (v_1^2 + v_2^2)
    \end{equation*}
    This is a relative degree 1 barrier function, which can be shown by
    \begin{align*}
        \frac{\partial \tilde{h}^{12}}{\partial \vx}&=\begin{bmatrix}
        2(p_{x_1} - p_{x_2})\\
        2(p_{y_1} - p_{y_2})\\ 0\\-2\mu v_1\\
        -2(p_{x_1} - p_{x_2})\\
        -2(p_{y_1} - p_{y_2})\\ 0\\-2\mu v_2\\
        \mathbf{0}_{8\times1}
        \end{bmatrix}\\
        \implies \frac{\partial \tilde{h}^{12}}{\partial \vx}\T G &=  \begin{bmatrix}0 & -2\mu v_1 & 0 & -2\mu v_2 & \mathbf{0}_{1\times4} \end{bmatrix}
    \end{align*}
    The other terms needed to set up \eqref{eq:qp} can be calculated as:
    \begin{align*}
        \frac{\partial \tilde{h}^{12}}{\partial \vx}\T f &= 2(p_{x_1} - p_{x_2})v_1\cos(\theta_1) + 2(p_{y_1} - p_{y_2})v_1\sin(\theta_1)\\
        &- 2(p_{x_1} - p_{x_2})v_2\cos(\theta_2) - 2(p_{y_1} - p_{y_2})v_2\sin(\theta_2)\\
        \frac{1}{2}\tr\big( \frac{\partial^2 \tilde{h}^{12}}{\partial \vx^2} \Sigma\Sigma\T\big) &= -2\mu\sigma^2
    \end{align*}
    The QP can be set up with
    \begin{align*}
        C_{12} &= -\dfrac{\partial \tilde{h}^{12} }{\partial \vx}^{\text{T}}G\\
        d_{12} &= \alpha(\tilde{h}^{12}(\vx))+\dfrac{1}{2}\tr \big( \dfrac{\partial^2 \Tilde{h}^{12}}{\partial \vx^2}\Sigma\Sigma\T\big) + \frac{\partial \tilde{h}^{12}}{\partial \vx}^{\text{T}}f
    \end{align*}
\end{enumerate}
\par Regarding values of common hyper-parameters used to train the Safe FBSDE controllers for the 2D car problems, we used the following,
\begin{center}
 \begin{tabular}{||c| c| c||} 
 \hline
 \# & \textbf{Parameter Name} & \textbf{Parameter Value} \\  
 \hline\hline
 1 & batch size (task i.) & 128 \\ 
 \hline
 2 & batch size (task ii.) & 256 \\ 
 \hline
 3 & training iterations & 1000 \\
 \hline
 4 & $\sigma$ & 0.1 \\
 \hline
 5 & learning rate & $5e^{-3}$ \\
 \hline
 6 & number of neurons per LSTM layer & 16 \\
 \hline
 7 & optimizer & Adam \\
 \hline
\end{tabular}
\end{center}

For task (i.), we used $\mu=0.05$ and $\gamma = 1$, while for task (ii.) we used, $\mu=0.1$ and $\gamma=1$.

\begin{figure}[!tbp]
  \centering
  \begin{minipage}[b]{0.32\textwidth}
    \includegraphics[width=\textwidth]{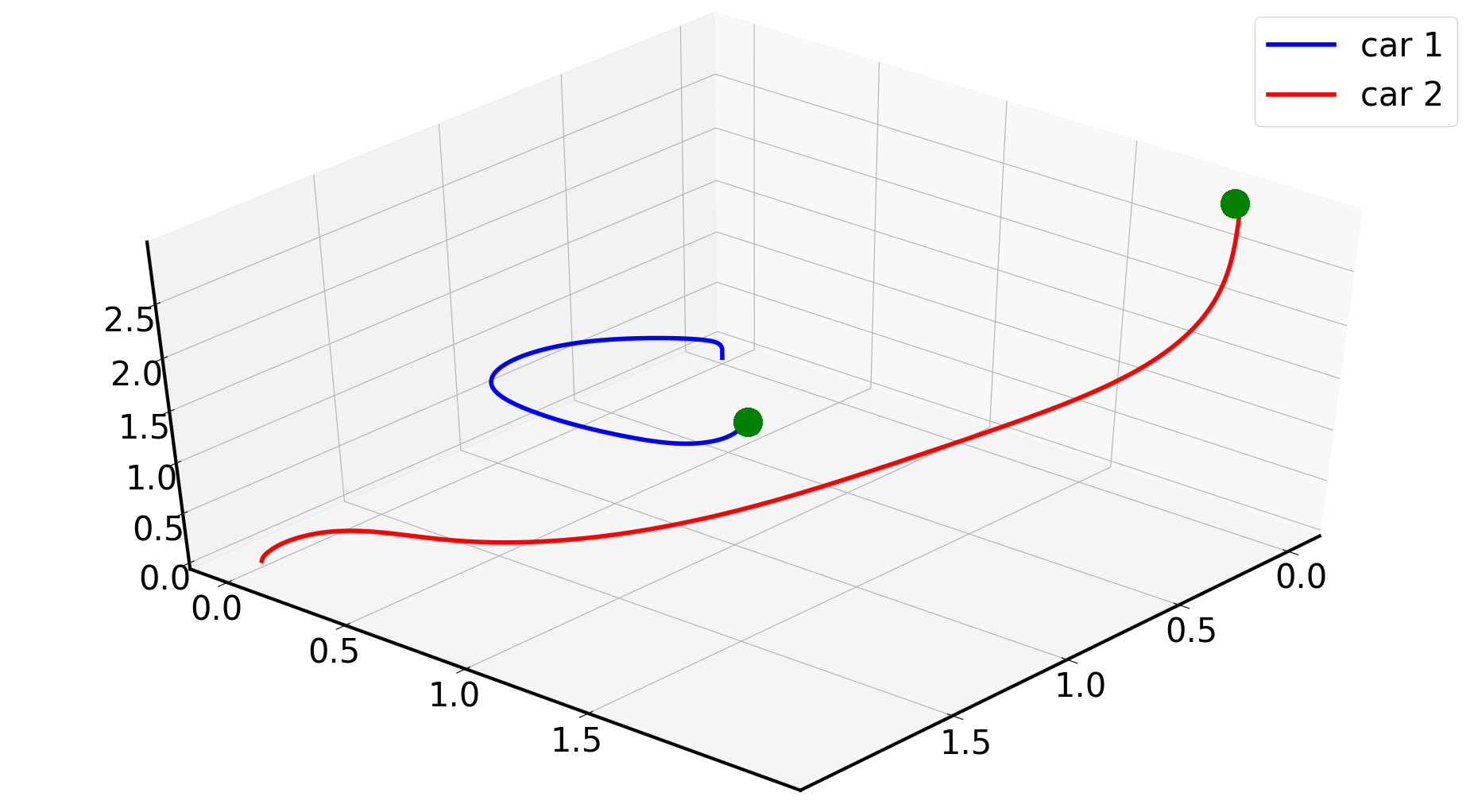}
  \end{minipage}
  \begin{minipage}[b]{0.32\textwidth}
    \includegraphics[width=\textwidth]{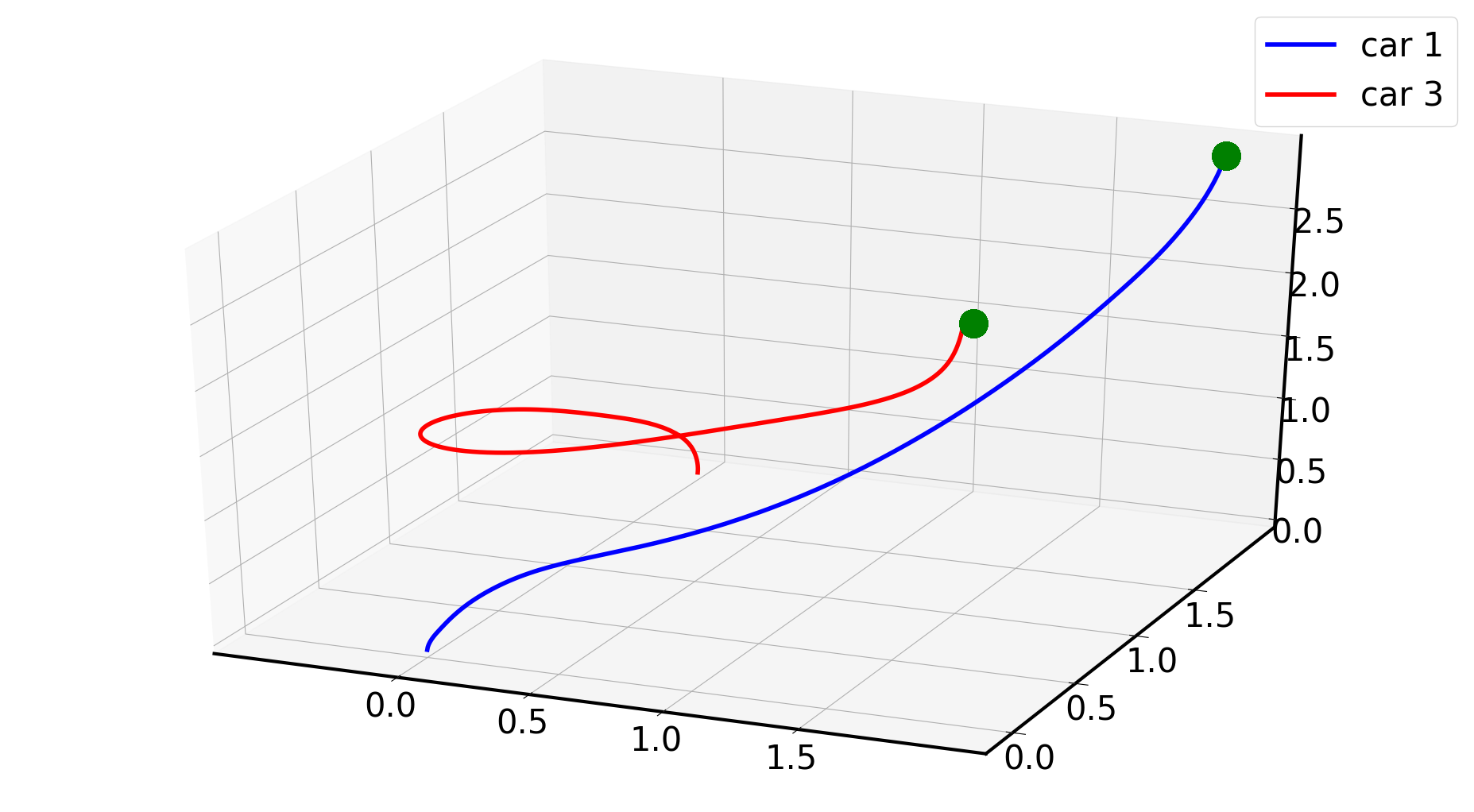}
  \end{minipage}
  \begin{minipage}[b]{0.32\textwidth}
    \includegraphics[width=\textwidth]{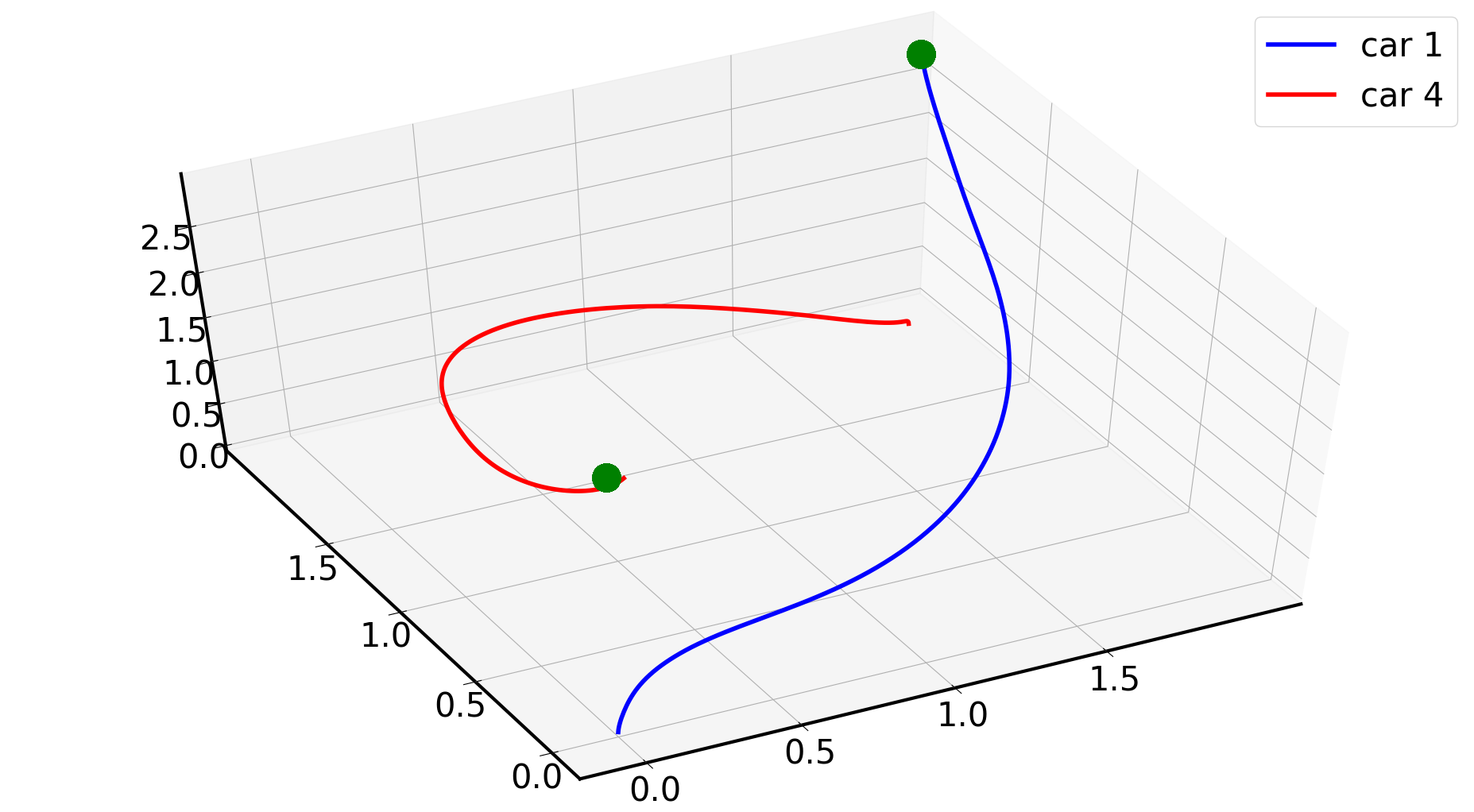}
  \end{minipage}
\end{figure}

\begin{figure}[!tbp]
  \centering
  \begin{minipage}[b]{0.32\textwidth}
    \includegraphics[width=\textwidth]{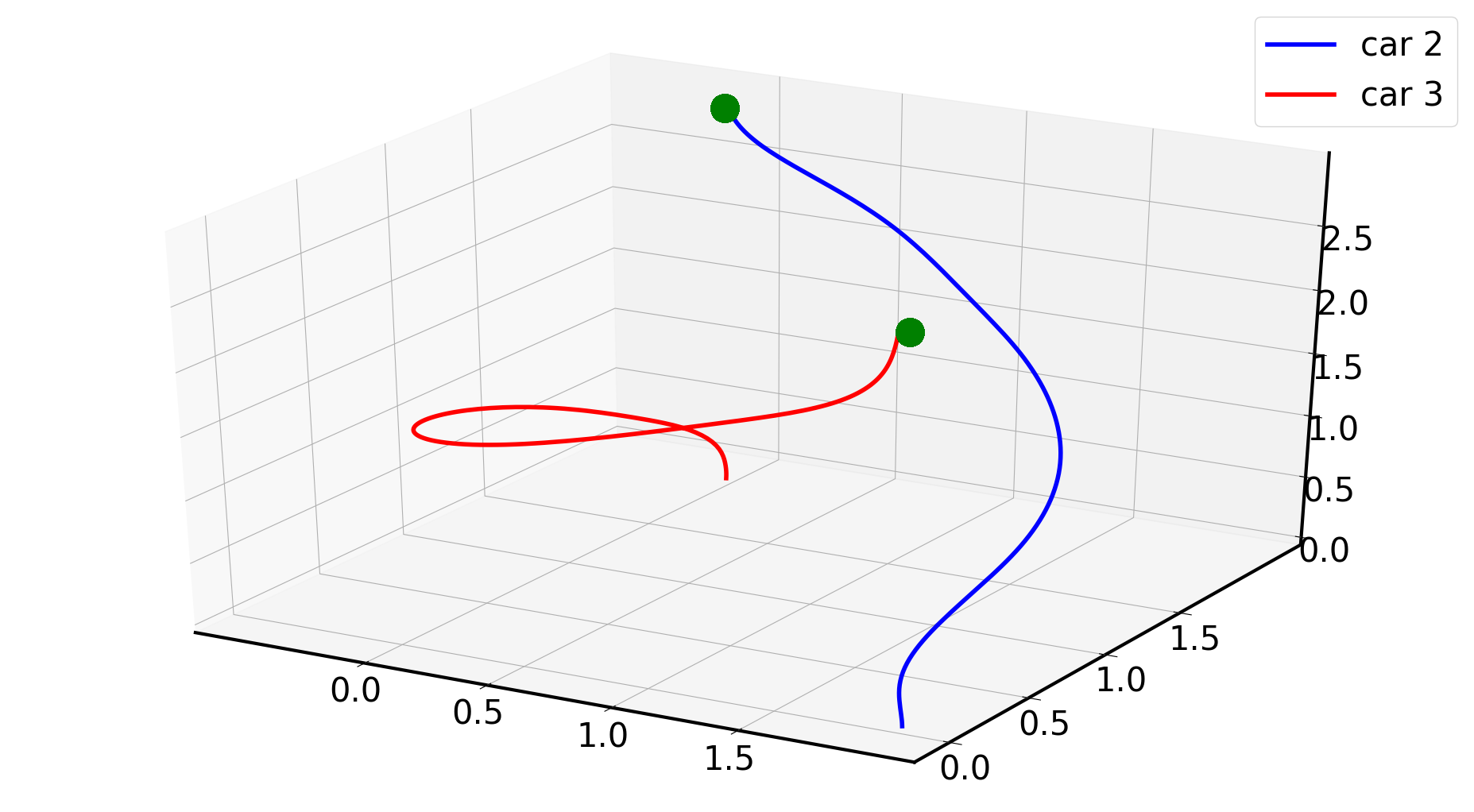}
  \end{minipage}
  \begin{minipage}[b]{0.32\textwidth}
    \includegraphics[width=\textwidth]{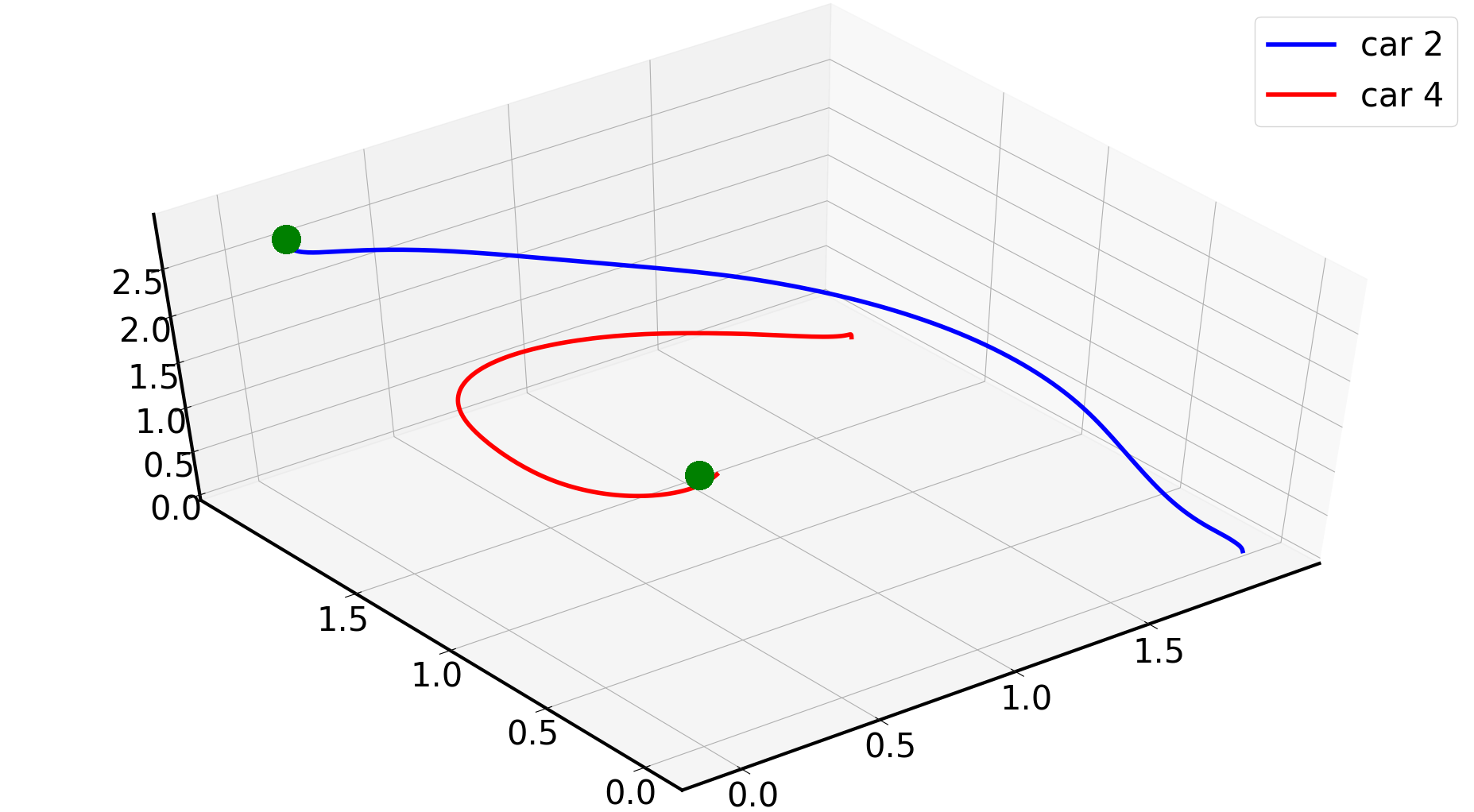}
  \end{minipage}
  \begin{minipage}[b]{0.32\textwidth}
    \includegraphics[width=\textwidth]{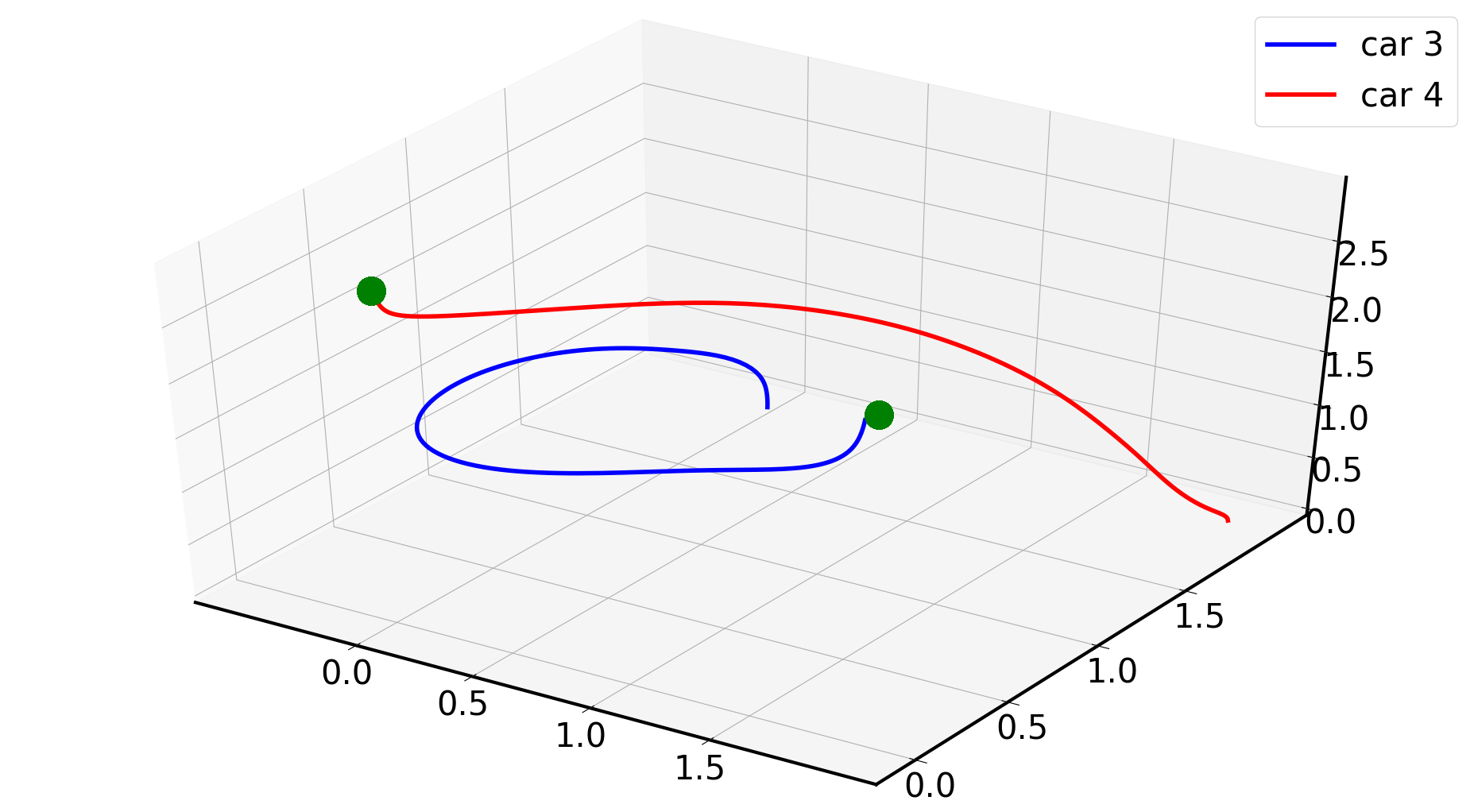}
  \end{minipage}
  \caption{\textbf{2D Car Collision Avoidance Average Testing Performance:} Figures above show successful collision avoidance performance of the trained policy averaged over 128 trials.  Vertical axes represent \textit{time (s)} and horizontal axes represent \textit{(x,y) position}.}
  \label{fig:2dcar_collision_test_3D}
\end{figure}

\subsection{Animation of Multi-Car Collision Avoidance}
Attached with this supplementary material is an animated video of the multi-car collision avoidance problem. For this animation we used the first batch element (from a batch of $128$) from the respective iterations seen in the video. However, when demonstrating performance during testing, we use average performance. As seen in the plot Fig.\ref{fig:2dcar_collision_test}, the variance of the trajectories is considerably small and therefore it suffices to show the mean performance. The mean cost is also the cost functional used in the problem formulation. The colored cicles indicate the diagonally opposite targets for each car. The dotted circles show the 2 car radii minimum allowable safe distance.    

\subsection{Algorithms}
The Alg.\ref{alg:FBSDEalgorithm} below, is very similar to work in \cite{pereira2019learning}, with the following changes: computation of the optimal control (Hamiltonian QP instead of closed form expression), propagation of the value function (no need for Girsanov's theorem), computation of an additional target for training the deep FBSDE network (i.e. the true gradient of the value function) and additional terms in the loss function. The colored circles in the video are the respective targets of each car. 

\begin{algorithm}[h!]
\caption{Finite Horizon Safe \ac{FBSDE} Controller}
    \begin{algorithmic}
          \STATE \textbf{Given:}\\ 
          $\vx_0=\xi,\;\vf,\; \vG$, $\vSigma$: Initial state and system dynamics;\\ $\phi,\; q, \; \vR$: Cost function parameters;\\ $a,b,c,d$: Loss function parameters;\\$N$: Task horizon, $K$: Number of iterations, $M$: Batch \\size; $\Delta t$: Time discretization;
          $\lambda$: weight-decay parameter;
          \STATE \textbf{Parameters:}\\
          $\psi=V(\vx_0,0)$: Value function at $t=0$;\\
          $\theta$: Weights of all Linear and LSTM layers;
          \STATE \textbf{Initialize neural network parameters};
          \STATE \textbf{Initialize (for every batch element):} $\{\vx^i_0\}_{i=1}^{M},\;\vx^i_0=\xi $;
          $\{\psi^i_0\}_{i=1}^{M},\;\psi^i_0=V(\vx^i_0,0)$
          \FOR{$k=1$ \TO $K$}
            \FOR{$i=1$ \TO $M$}
                \FOR{$t=1$ \TO $N-1$}
                    \STATE Predict value function gradient:\\ $V_{\vx,t}^i=f_{FBSDE}(\vx_t^i;\theta^k)$
                    \STATE Solve constrained Hamiltonian minimization quadratic program:\\ $\bar{\vu}^{*i}_t=f_{safe}(\vx_t^i, V_{\vx,t}^i)$
                    \STATE Sample Brownian noise: \\$\Delta \vw^i_t \sim \mathcal{N}(0, \vSigma\Delta t)$
                    \STATE Propagate value function: $V^i_{t+1} = V^i_t - \big(q(\vx_t)+\frac{1}{2}{\bar{\vu}^{*i\text{T}}_t} R\bar{\vu}^{*i}_t\big) \Delta t + V_{\vx,t}^{i\mathrm{T}}\vSigma_t^i \Delta \vw^i_t$
                    \STATE Propagate system state: $\vx^i_{t+1}=\vx^i_t + \vf_t^i \Delta t + \vG_t^i \bar{\vu}^{*i}_t\Delta t +\vSigma_t^i\Delta \vw^i_t $
                \ENDFOR
                \STATE Compute targets: true terminal value and true value function gradient,\\ $V^{*i}_N=\phi\big(\vx^i_N\big)$; $V_{\vx,N}^{*i}=\phi_\vx\big(\vx_N^i\big)$
            \ENDFOR
            \STATE Compute mini-batch loss: \\$\mathcal{L}= \displaystyle \frac{1}{M} \sum_{i=1}^M a\,\| V^{*i}_N - V^i_N\|^2_2+ b\,\| V^{*i}_{\vx,N} - V^i_{\vx,N}\|^2_2 + c\,\|V^{*i}_N\|^2_2 + d\,\|V^{*i}_{\vx,N}\|^2_2 + \lambda\, \| \theta^k \|^2_2 $
            \STATE Gradient step:\\ $\theta^{k+1}, \psi^{k+1}\leftarrow$ Adam.step($\mathcal{L}, \theta^k, \psi{k}$)       
          \ENDFOR
          \RETURN $\theta^K, \psi^K$
    \end{algorithmic}
    \label{alg:FBSDEalgorithm}
\end{algorithm}

\begin{algorithm}[h!]
\caption{Stochastic \ac{CBF} layer (Safe OptNet layer)}
    \begin{algorithmic}
    \STATE \textbf{Given:}\\
    $h$: Safe set characterizing barrier function; $\alpha(h(\vx)) =\gamma h(\vx)$;\\
    $\vf, \vG, \vSigma$: System dynamics; $L$: Maximum QP iterations, $R$: control cost positive definite matrix
    \STATE \textbf{Input:} Current system state ($\vx_t$) and current predicted value function gradient ($V_{\vx,t}$)\\
    \STATE \textbf{Set up the \ac{QP} problem:}\\
    $Q=R,\, q=V_{\vx}^\mathrm{T}G, \,C=-\dfrac{\partial h}{\partial \vx}\T \vG(\vx),\, d=\alpha(h(\vx))+\dfrac{\partial h}{\partial \vx}\T \vf(\vx) + \dfrac{1}{2}\tr \bigg(\vSigma(\vx)\T\dfrac{\partial^2 h}{\partial \vx^2}\vSigma(\vx)\bigg)$
    \STATE \textbf{Initialize solution:}\\
    initialize feasible $\vu$, slack variable: $s\in\Rb_+^1$ and Lagrange multiplier for \ac{CBF} inequality constraint: $\lambda \in\Rb_+$ and  Residue Tolerance: $\epsilon$ 
    \FOR{$j=1$ to $L$}
     \STATE  $(\Delta \vu_j^{aff}, \Delta s_j^{aff}, \Delta \lambda_j^{aff}) \leftarrow $Affine Scale $(\vu_j,s_j,\lambda_j)$
     \STATE $(\Delta \vu_j^{cc},\Delta s_j^{cc}, \Delta \lambda_j^{cc})\leftarrow$ Centering-Corrector $(\vu_j,s_j,\lambda_j)$
     \STATE $\zeta\leftarrow$ Residual$(\vu_j,s_j,\lambda_j)$
     \STATE $\alpha\leftarrow$ Step Size$(\vu_j,s_j,\lambda_j)$ 
      
    \STATE $\vu_{(j+1)}= \vu_j+\alpha (\Delta \vu_j^{aff}+\Delta \vu_j^{cc})$\\ $s_{(j+1)}= s_j+\alpha (\Delta s_j^{aff}+\Delta s_j^{cc})$ \\
    $\lambda_{(j+1)}= \lambda_j+\alpha(\Delta \lambda_j^{aff}+\Delta \lambda_j^{cc})$\\
    \IF{$\zeta\leq \epsilon$}
    \STATE Break
    \ENDIF
    
    \ENDFOR\\
    Return $\bar{\vu}^*\leftarrow$ $\vu_j$
    \end{algorithmic}
    \label{alg:qp}
\end{algorithm}
In Alg.\ref{alg:qp}, for details regarding \textbf{Affine Scale}, \textbf{Centering-Corrector}, \textbf{Residul} and \textbf{Step-Size} please see the primal-dual interior point method detailed in \cite{nocedal2006numerical} or refer to code provided on github in \cite{amos2017optnet}.

\end{document}